\documentclass[12pt]{article}
\usepackage{amsmath,amssymb}
\usepackage{graphicx,psfrag,epsf}
\usepackage{enumerate}
\usepackage{natbib}
\usepackage{url} 
\usepackage{color}
\usepackage{caption}
\usepackage{subcaption}
\usepackage[colorlinks=true,linkcolor=blue]{hyperref}
\usepackage{algorithm}
\usepackage{algpseudocode}
\usepackage{multirow}
\usepackage{mathtools}
\usepackage{makecell}

\newcommand{\blind}{1}
\usepackage{manyfoot}
\DeclareNewFootnote{A}[arabic]

\begin{document}

\def\spacingset#1{\renewcommand{\baselinestretch}%
{#1}\small\normalsize} \spacingset{1}


\if1\blind
{
  \title{\bf Conglomerate Multi-Fidelity Gaussian Process Modeling, with Application to Heavy-Ion Collisions}
    \author{
    Yi Ji\footnotemarkA[1]\protect\phantom{\footnotesize 1}\textsuperscript{,}\footnotemarkA[5],\quad Henry Shaowu Yuchi\footnotemarkA[2]\protect\phantom{\footnotesize 1}\textsuperscript{,}\footnotemarkA[5]\\
    Derek Soeder\footnotemarkA[3],\quad 
    J.-F. Paquet\footnotemarkA[3]\protect\phantom{\footnotesize 1}\textsuperscript{,}\footnotemarkA[4]\\
    Steffen A. Bass\footnotemarkA[3],\quad 
    V. Roshan Joseph\footnotemarkA[2],\quad 
    C. F. Jeff Wu\footnotemarkA[2], \quad Simon Mak\footnotemarkA[1]\protect\phantom{\footnotesize 1}\textsuperscript{,}\footnote{Corresponding author}
    }
    \footnotetext[1]{Department of Statistical Science, Duke University}
    \footnotetext[2]{H. Milton Stewart School of Industrial \& Systems Engineering, Georgia Institute of Technology}
    \footnotetext[3]{Department of Physics, Duke University}
    \footnotetext[4]{Department of Physics and Astronomy \& Department of Mathematics, Vanderbilt University}
    \footnotetext[5]{Joint first authors}
  \maketitle
} \fi

\if0\blind
{
  \bigskip
  \bigskip
  \bigskip
  \begin{center}
    {\LARGE\bf Conglomerate Multi-Fidelity Gaussian Process Modeling, with Application to Heavy-Ion Collisions}
\end{center}
  \medskip
} \fi
\vspace{-0.4in}
\bigskip

\begin{abstract}
In an era where scientific experimentation is often costly, multi-fidelity emulation provides a powerful tool for predictive scientific computing. While there has been notable work on multi-fidelity modeling, existing models do not incorporate an important ``conglomerate'' property of multi-fidelity simulators, where the accuracies of different simulator components are controlled by different fidelity parameters. Such conglomerate simulators are widely encountered in complex nuclear physics and astrophysics applications. We thus propose a new CONglomerate multi-FIdelity Gaussian process (CONFIG) model, which embeds this conglomerate structure within a novel non-stationary covariance function. We show that the proposed CONFIG model can capture prior knowledge on the numerical convergence of conglomerate simulators, which allows for cost-efficient emulation of multi-fidelity systems. We demonstrate the improved predictive performance of CONFIG over state-of-the-art models in a suite of numerical experiments and two applications, the first for emulation of cantilever beam deflection and the second for emulating the evolution of the quark-gluon plasma, which was theorized to have filled the Universe shortly after the Big Bang.
\end{abstract}

\noindent%
\vfill

\newpage
\spacingset{1.45}

\section{Introduction}
\label{sec:intro}
Computer experimentation is widely used for modeling complex scientific and engineering systems, particularly when physical experiments are costly, unethical, or impossible to perform. This shift from physical to computer experimentation has found success in a wide range of physical science applications, including rocket design \citep{mak2018efficient}, solar irradiance modeling \citep{sun2019synthesizing} and 3D printing \citep{chen2021function}. However, as systems become more complex and realistic, such computer experiments also become more expensive, thus placing a heavy computational burden on design exploration and optimization. Statistical \textit{emulators} \citep{santner2003design} have shown great promise in tackling this limitation. The idea is simple but effective: computer experiments are first performed at carefully chosen design points, then used as training data to fit an \textit{emulator} model to efficiently predict and quantify uncertainty on the expensive virtual experiment.


In recent years, however, with the increasing sophistication of modern scientific problems, an emerging challenge for emulators is the simulation of high-fidelity training data, which can be prohibitively expensive. One way to address this is via \textit{multi-fidelity emulation}, which makes use of training simulation data of multiple \textit{fidelities} (or accuracy) for model fitting. Such multi-fidelity data can often be generated by varying different \textit{fidelity parameters}, which control the precision of the numerical experiment. There are a wide variety of fidelity parameters, ranging from mesh sizes for finite element analysis \citep{park1997determination,more2015effect} to time-steps for dynamical system simulation \citep{vanden2003numerical}. The goal is to leverage information from lower-fidelity (but cheaper) simulations to enhance predictions for the high-fidelity (but expensive) model, thus allowing for improved emulation and uncertainty quantification (for the highest-fidelity code) at lower computational costs.



There has been much recent work on multi-fidelity emulation, particularly for Gaussian process (GP) modeling. This includes the seminal work of \cite{kennedy2000predicting}, which presented a first-order autoregressive model for integrating information over a hierarchy of simulation models, from lowest to highest fidelity. This Kennedy-O'Hagan model has then been extended in various works, including a Bayesian hierarchical implementation in \cite{qian2006building}, the multi-fidelity optimization in \cite{forrester2007multi}, and the nonlinear fusion model in \cite{perdikaris2017nonlinear}. \cite{tuo2014surrogate} proposed a multi-fidelity emulator for finite element analysis (FEA), which utilizes the discretization mesh size as the single fidelity parameter. This emulator models the bias induced by discretization mesh elements via a GP, and is related to the state-of-the-art grid convergence index approach typically employed in FEA \cite{roache1994perspective, bect2021quantification}. Such multi-fidelity models have been widely applied in engineering design and scientific computing; see, e.g., \cite{kou2019multi,shi2020multi,jin2021combining,liyanage2022efficient}. Experimental design for such emulators have been explored \citep{xiong2013sequential}, including a sequential design strategy in \cite{he2017optimization}. Similar ideas have also been applied for broader applications in data fusion \citep{ghoreishi2019multi}, Bayesian optimization \citep{poloczek2017multi, moss2021gibbon}, and transfer learning \citep{tighineanu2022transfer}.

\begin{figure}[!t]
    \centering
    \includegraphics[width=1\textwidth]{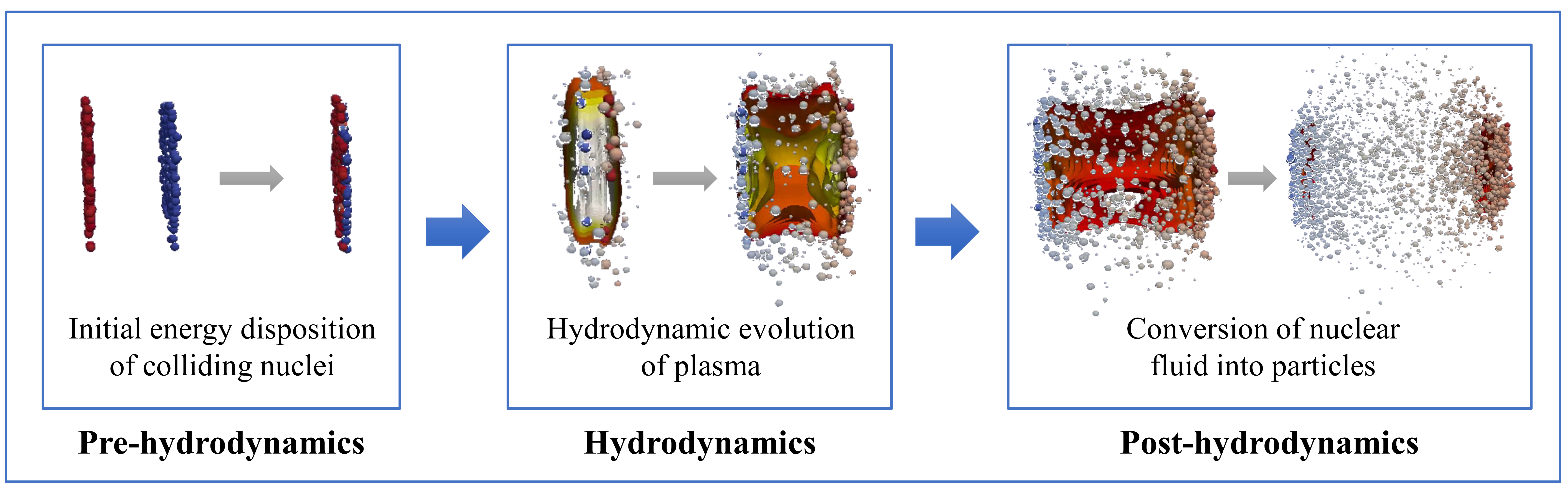}
    \caption{Three-stage simulation of the quark-gluon plasma.}
    \label{fig:QGP_intro}
\end{figure}

The above methods, however, have limitations when applied to our motivating nuclear physics application. Here, we are studying the Quark-Gluon Plasma (QGP), a deconfined phase of nuclear matter consisting of elementary quarks and gluons. The QGP was theorized to have filled the Universe shortly after the Big Bang, and the study of this plasma sheds light on the properties of this unique phase of matter. This plasma can be simulated at a small scale by virtually colliding heavy ions together at near-light speeds in particle colliders. Simulating such collisions requires a ``conglomerate'' system of complex dynamical models to faithfully capture the detailed evolution of the plasma. Consider in particular the three-stage simulation framework in \cite{everett2021multisystem} (see also~\citep{gale2013hydrodynamic,heinz2013collective,de2016hydrodynamic}), which models the initial energy disposition of the heavy ions, the hydrodynamic evolution of the plasma after the collision, and the subsequent conversion of nuclear fluid into particles. Figure~\ref{fig:QGP_intro} visualizes this conglomerate (specifically, multi-stage) procedure. At each stage, the simulation of the component physics can involve \textit{multiple} and \textit{different} fidelity parameters, controlling, e.g., the size of the hydrodynamics spatial mesh, or the time-scale for dynamic evolution.





This \textit{conglomerate multi-fidelity} framework, where the simulator comprises multiple submodels for simulating different physics of a complex phenomenon, poses several challenges for existing multi-fidelity emulators. First, since there are multiple fidelity parameters to set for each simulation stage, the resulting simulation runs typically cannot be ranked from lowest to highest fidelity, which is required for a direct application of Kennedy-O'Hagan-type models. For example, to gauge the effects of three fidelity parameters, the physicist may choose to run the simulator in three different ways, each with higher fidelity at one stage and lower fidelity at the remaining stages. A priori, it is unclear if these three simulation approaches can be ranked from lowest to highest fidelity. Second, unlike the multi-fidelity emulator in \cite{tuo2014surrogate} (which allows only one fidelity parameter), there are \textit{multiple} fidelity parameters that should be accounted for when training emulators with conglomerate simulations. Neglecting this conglomerate structure for emulation can result in significantly poorer predictive performance, as we show later. A broader emulation model is thus needed to tackle the challenges presented by conglomerate multi-fidelity simulators, which are widely encountered in nuclear physics \citep{ji2021graphical} and astrophysics \citep{ho2022multifidelity}. 




We propose in this work a new GP emulator that addresses these challenges. The proposed CONglomerate multi-FIdelity Gaussian process (CONFIG) model makes use of a novel \textit{non-stationary} covariance function, which captures prior information on the numerical convergence of conglomerate simulators. Our emulator is applicable for a variety of complex multi-physics simulators, where each physics (with its corresponding fidelity parameters) is jointly simulated via a conglomerate framework. By embedding this underlying conglomerate structure within its kernel specification, the CONFIG model can yield improved emulation performance and uncertainty quantification over existing methods for predicting the limiting highest-fidelity simulator. This is demonstrated in a suite of numerical experiments, a beam deflection problem in finite element analysis, and an application to the motivating heavy-ion collision problem. Section \ref{sec:lit} reviews several existing multi-fidelity emulators and outlines the motivating QGP problem. Section \ref{sec:method} presents the model specification for the proposed CONFIG emulator. Section \ref{sec:imp} discusses implementation details for CONFIG, including parameter estimation and experimental design. Section \ref{sec:numerical} compares the proposed model with existing methods on a suite of numerical experiments. Finally, Section \ref{sec:apply} demonstrates the effectiveness of CONFIG for the motivating QGP application as well as a cantilever beam deflection problem. Section \ref{sec:conclusion} concludes the paper.

\section{Preliminaries \& Motivation}
\label{sec:lit}


\begin{figure}[!t]
    \centering
    \includegraphics[width=.8\textwidth]{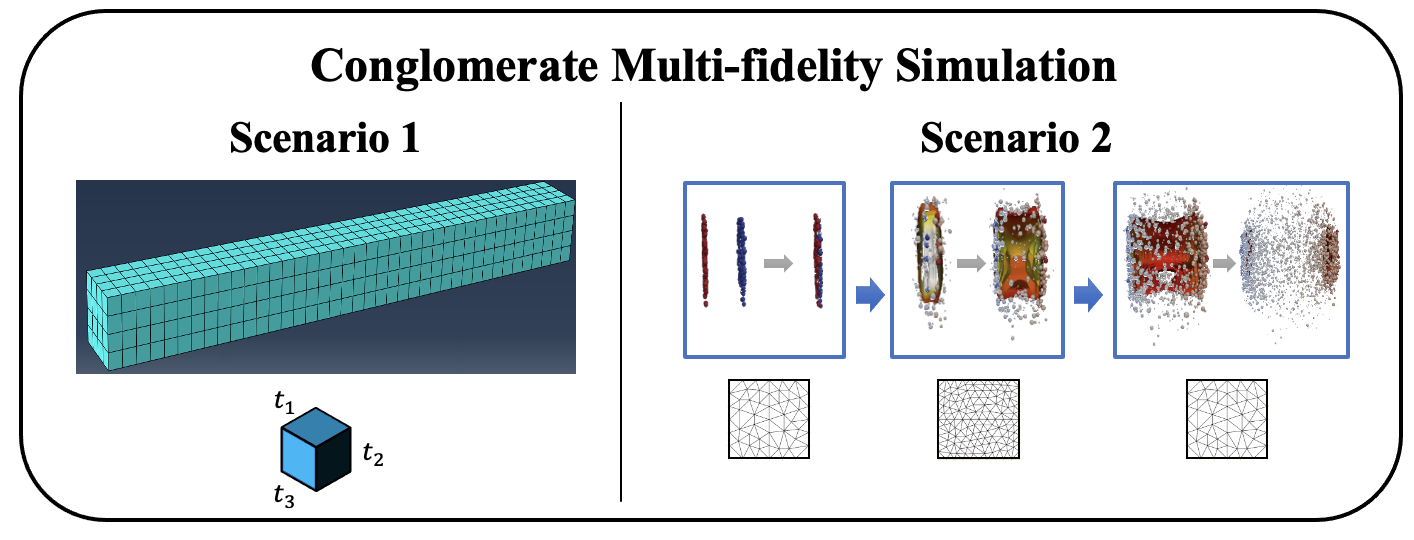}
    \caption{Visualizing examples of Scenarios 1 and 2 for conglomerate multi-fidelity simulations. The left plot shows an example of Scenario 1 for cantilever beam deflection, where the three fidelity parameters specify the size of the finite elements for the beam. The right plot shows an example of Scenario 2 for heavy-ion collisions, where different fidelity parameters control simulation precision at different stages of the collision system.}
    \label{fig:overview}
\end{figure}

In this section, we first provide an overview of conglomerate multi-fidelity simulators and their use for complex multi-physics applications. We then briefly introduce the Gaussian process model and review the Kennedy-O'Hagan model in \citep{kennedy2000predicting} and the multi-fidelity model in \cite{tuo2014surrogate}. Finally, we discuss the limitations of such models for our QGP application, thus motivating the proposed CONFIG model.


\subsection{Conglomerate Multi-fidelity Simulation}

\label{sec:cong}


With an urgent need for reliable simulation of complex phenomena involving multiple physical mechanisms and/or components, conglomerate multi-fidelity simulations are now increasingly used in modern scientific and engineering applications, such as structural studies \cite{sun2010two,diazdelao2012bayesian}, engine combustion \citep{narayanan2023physics} and high-energy physics \cite{kumar2023inclusive}. Such simulators model the complex phenomenon via either multiple submodels that account for different physics (e.g., hydrodynamic evolution, nuclear particlization), or multiple components (e.g., spatial mesh, time discretization) that facilitate the simulation procedure. Consequently, the simulation of the \textit{overall} phenomenon typically involves \textit{multiple} fidelity parameters, each controlling the simulation accuracy of individual submodels or components. This poses a key challenge for existing emulator models.


To tackle this, it is useful to first understand different types of conglomerate simulators encountered in applications. In our experience, this falls roughly into two scenarios (see Figure \ref{fig:overview}):
\begin{itemize}
    \item \textbf{Scenario 1}: The simulator consists of multiple fidelity parameters for simulating a \textit{single} mechanism or phenomenon. Such parameters control different means for varying simulation precision, e.g., via spatial meshing or temporal discretization. One example is the FEA of a cantilever beam deflection under stress, where three mesh fidelity parameters can be used used for each dimension of the three-dimensional finite element analysis. We will investigate this application further in Section \ref{sec:beam}.
    \item \textbf{Scenario 2}: The simulator comprises multiple stages that are performed \textit{sequentially} over time, where a separate phenomenon is simulated at each stage, with associated fidelity parameters. \textit{Multiple} mechanisms are thus involved in simulating the desired phenomenon. This is the case for our motivating nuclear physics problem (Figure \ref{fig:QGP_intro}), where multiple mesh size parameters control simulation precision in each of the three consecutive stages for heavy-ion collisions. We will investigate this application further in Section \ref{sec:multistage}.
\end{itemize}
Motivated by these two scenarios, we will present later two variations of the CONFIG model that tackle each of these scenarios; more on this in Section \ref{sec:method}.



\subsection{Gaussian Process Modeling}

Gaussian process (GP) modeling is a popular Bayesian nonparametric approach for supervised learning \citep{williams2006gaussian}, with broad applications for computer experiments \citep{santner2003design}. The specification of a GP model involves two key ingredients: the mean function and the covariance function. Let $\mathbf{x} \in [0,1]^p$ be the input parameters (sufficiently scaled) for the simulator, and let $\eta(\mathbf{x})$ be the corresponding output of the simulator. A GP model places the following prior on the unknown response surface $\eta(\cdot)$:
\begin{equation}
    \eta(\cdot)\sim \mathcal{GP}(\mu(\cdot),k(\cdot,\cdot)).
\end{equation}
Here, $\mu(\cdot)$ is the mean function controlling the centrality of the stochastic process. If appropriate basis functions $\boldsymbol{f}(\boldsymbol{x})$ are known, one can model the mean function as $\mu(\boldsymbol{x})=\boldsymbol{f}(\boldsymbol{x})^T\boldsymbol{\beta}$, where $\boldsymbol{\beta}$ are the corresponding coefficients on $\boldsymbol{f}(\boldsymbol{x})$. In the absence of such information, $\mu(\cdot)$ is typically set to be a constant. The function $k(\cdot,\cdot)$ is the covariance function that controls the smoothness of its sample paths. Common choices of $k(\cdot,\cdot)$ include the squared-exponential and Mat\'ern kernels \citep{santner2003design}.

Let $\mathcal{D}=\{\mathbf{x}_1,\cdots,\mathbf{x}_n\}$ denote the simulated input points, and $\mathbf{y}=[\eta(\mathbf{x}_1),\cdots,\eta(\mathbf{x}_n)]$ be the simulated outputs. Assuming that the kernel hyperparameters are fixed and known (we will discuss the estimation of such parameters later in Section \ref{sec:paraminf}), the predictive distribution $\eta(\mathbf{x}^*)$ at the new input $\mathbf{x}^*$ conditional on data $\{\mathcal{D},\mathbf{y}\}$ is given by:
\begin{equation}
\eta(\mathbf{x}^*)|\mathcal{D},\mathbf{y}\sim \mathcal{GP}(\hat{\mu}(\mathbf{x}^*),s^2(\mathbf{x}^*)).
\label{eq:gppred}
\end{equation}
Here, the posterior mean and variance are given by:
\begin{align}
\begin{split}
\hat{\mu}(\mathbf{x}^*) &= \mu(\mathbf{x}^*) + \mathbf{k}(\mathbf{x}^*,\mathcal{D})^T\mathbf{K}(\mathcal{D})^{-1}(\mathbf{y}-\boldsymbol{\mu}(\mathcal{D})),\\
s^2(\mathbf{x}^*) &= k(\mathbf{x}^*,\mathbf{x}^*)-\mathbf{k}(\mathbf{x}^*,\mathcal{D})^T\mathbf{K}(\mathcal{D})^{-1}\mathbf{k}(\mathbf{x}^*,\mathcal{D}),
\end{split}
\label{eq:predeqn}
\end{align}
where $\mathbf{k}(\mathbf{x}^*,\mathcal{D})=[k(\mathbf{x}^*,\mathbf{x}_1),\cdots,k(\mathbf{x}^*,\mathbf{x}_n)]$ is the vector of covariances, $\boldsymbol{\mu}(\mathcal{D}) = [\mu(\mathbf{x}_1), \cdots,\allowbreak \mu(\mathbf{x}_n)]$ is the vector of means, and $\mathbf{K}(\mathcal{D})$ is the covariance matrix for the training data. The models introduced later in this paper will make use of these closed-form predictive equations with different choices of covariance functions.

\subsection{The Kennedy-O'Hagan model}
\label{sec:KOH}
In the seminal work of \cite{kennedy2000predicting}, the authors proposed a first-order autoregressive model for linking outputs from a hierarchy of $H$ simulators, from lowest fidelity (simulator $1$) to the highest fidelity (simulator $H$). Let {$\eta_h(\mathbf{x})$ denote the output from simulator $h$} at standardized input parameters $\mathbf{x} \in [0,1]^p$. The Kennedy-O'Hagan (KOH) model is specified as:
\begin{equation}
    \eta_h(\mathbf{x})=\rho_{h-1}\eta_{h-1}(\mathbf{x})+\delta_h(\mathbf{x}), \quad h = 2, \cdots, H.
    \label{eq:KOH}
\end{equation}
Here, $\rho_{h-1}$ is a regression scale factor, and $\delta_h(\mathbf{x})$ is a bias term that models the discrepancy between simulator $h-1$ and $h$. The bias term $\delta_h(\mathbf{x})$ may be modeled by a stationary GP with a squared-exponential covariance function \citep{santner2003design}:
\begin{equation}
    \text{Cov}\left[\delta_h(\mathbf{x}),\delta_h(\mathbf{x}')\right] = \sigma_h^2 \text{ exp}\left\{ - \sum_{i=1}^p \phi_{h,i} (\mathbf{x}_i-\mathbf{x}'_i)^2\right\},
    \label{eq:sqr-exp}
\end{equation}
where $\phi_{h,i}$ is the weight parameter for the $i^{\text{th}}$ input parameter at the $h^{\text{th}}$ fidelity level. Such a model allows one to integrate information from a sequence of simulator models with varying fidelity levels, to efficiently emulate the highest-fidelity simulator model.

The KOH multi-fidelity model has subsequently been extended in a variety of ways, including a Bayesian implementation in \cite{qian2008bayesian} and a nonlinear extension in \cite{perdikaris2015multi}; see also \cite{reese2004integrated,diazdelao2012bayesian,fricker2013multivariate}. This modeling framework is also closely related to the idea of co-kriging \citep{stein1991universal} in spatial statistics, and was employed for sequential co-kriging design \citep{le2015cokriging}. However, the aforementioned methods assume that the multi-fidelity training data can be \textit{ranked} from lowest to highest fidelity. As such, this body of literature does not directly apply to the motivating problem of conglomerate multi-fidelity emulation, where simulation accuracy is controlled by \textit{multiple} fidelity parameters, and thus there is no clear ranking of training data from lowest to highest fidelity. There are several ways to force existing models on this problem, but each has its shortcomings. One could design the data such that the training simulations are ranked (e.g., increasing all fidelity parameters simultaneously), but this would result in highly inefficient designs which fail to sufficiently explore the space of fidelity parameters. One could also arbitrarily assign a \textit{single} ``artificial'' fidelity level for each simulation, which imposes a ranking on the training runs. This, however, \textit{neglects} the rich conglomerate multi-fidelity framework (i.e., the ``science'') for the simulator, which can lead to significantly poorer predictive performance from the emulator, as we show later.

\subsection{The Tuo-Wu-Yu model}
\label{sec:TWY}
For problems where the fidelity level is controlled by a \textit{single} continuous fidelity parameter $t$ (e.g., mesh size), \cite{tuo2014surrogate} proposed an alternate model (we call this the TWY model) which can make use of such information. Let $\eta(\mathbf{x},t)$ denote the deterministic code output at standardized inputs $\mathbf{x} \in [0,1]^p$ and at fidelity parameter $t$. Here, $t$ is typically assumed to be between 0 and 1, with a smaller $t$ indicating a finer mesh size or, equivalently, higher mesh density. The TWY model adopts the following model for $\eta(\mathbf{x},t)$:
\begin{equation}
    \eta(\mathbf{x},t)=\eta(\mathbf{x},0)+\delta(\mathbf{x},t)=:\phi(\mathbf{x})+\delta(\mathbf{x},t).
    \label{eq:tuo}
\end{equation}
Here, $\phi(\mathbf{x}) := \eta(\mathbf{x},0)$ denotes the ``exact'' simulation output at input $\mathbf{x}$ at the highest (limiting) fidelity $t=0$, and $\delta(\mathbf{x},t)$ denotes the discrepancy (or bias) between this exact solution and realized simulation output with mesh size $t$. In practical problems, the exact solution $\eta(\mathbf{x},0)$ is typically \textit{not obtainable} numerically since some level of approximation (e.g., mesh or time discretization) is needed for simulating the system. The goal is to leverage simulation training data of the form $\{\eta(\mathbf{x}_i,t_i)\}_{i=1}^n$ along with an appropriate model on \eqref{eq:tuo} to predict the exact solution $\eta(\mathbf{x},0)$.

Since $\phi(\mathbf{x})$ and $\delta(\mathbf{x},t)$ are unknown \textit{a priori}, these terms are modeled in \cite{tuo2014surrogate} by two independent Gaussian processes. For $\phi(\mathbf{x})$, a standard GP prior is assigned with constant mean and a stationary correlation (e.g., squared-exponential) function. For the bias term $\delta(\mathbf{x},t)$, a \textit{non-stationary} zero-mean GP prior is assigned, with covariance function:
\begin{equation}  
\begin{split}
    \text{Cov}[\delta(\mathbf{x_1},t_1),\delta(\mathbf{x_2},t_2)]&= \sigma_2^2 K_{\mathbf{x}}^\delta(\mathbf{x}_1,\mathbf{x}_2)\min (t_1,t_2)^l,
\end{split}
\label{eq:nonstationary}
\end{equation}
where $K_{\mathbf{x}}^\delta(\cdot,\cdot)$ is a stationary correlation function on input parameters $\mathbf{x}$. One can view this as a product of two kernels, where the kernel on the fidelity parameter $t$ is non-stationary and closely resembles that of a Brownian motion. This separable kernel structure has been utilized in \cite{picheny2013nonstationary} as well.

This choice of non-stationary kernel over the single fidelity parameter $t$ can be reasoned from a Bayesian modeling perspective. Consider the GP model with covariance function \eqref{eq:nonstationary} as a prior model on discrepancy $\delta(\mathbf{x},t)$. Before observing the data, one can show that:
\begin{equation}
\lim_{t \rightarrow 0} \delta(\mathbf{x},t) = 0, \quad \text{for all $\mathbf{x} \in [0,1]^p$ almost surely.}
\label{eq:limittwy}
\end{equation}
The TWY model thus assumes a priori that the discrepancy term should converge to 0 as fidelity parameter $t$ goes to 0, or equivalently, the simulation output $\eta(\mathbf{x},t)$ converges to the exact solution $\phi(\mathbf{x})$, as we increase the fidelity of the simulator. This can be seen as a way of integrating \textit{prior} information on the numerical convergence of the simulator within the prior specification of the emulator model. One can further set the kernel parameter $l$ to capture additional information on known numerical convergence rates of the simulator; see \cite{tuo2014surrogate} for details.


For the target conglomerate setting where \textit{multiple} fidelity parameters are present, the TWY model needs to be further extended. A simple modification might be to first assign for each simulation run an ``artificial'' fidelity, e.g., the average of the multiple fidelity parameters, then use this single aggregate fidelity level with the TWY model for multi-fidelity emulation. However, such an approach ignores the rich conglomerate structure of the simulation framework, which can lead to poor predictive performance. We show later that, by integrating directly the conglomerate multi-fidelity nature of the simulation framework (i.e., the ``science'') within the CONFIG model, we can achieve significantly improved predictive performance in numerical experiments and for the motivating nuclear physics application.

\section{The CONFIG Model}
\label{sec:method}
Given these limitations, we now present the proposed CONFIG model for the efficient emulation of conglomerate multi-fidelity simulations. Our model adopts a novel non-stationary Gaussian process model which captures \textit{prior} information on the numerical convergence behavior of \textit{conglomerate} simulators. Below, we outline the general CONFIG model specification, then present two choices of non-stationary covariance functions which capture this desired prior information. 

Let $\mathbf{x}\in[0,1]^p$ be the vector of $p$ standardized simulation inputs for the computer code (again assumed to be deterministic), and suppose there are $q$ fidelity parameters (denoted by $\mathbf{t}\in[0,1]^q$) which control simulation accuracy in the code. These may, e.g., consist of different mesh sizes for domain discretization and time steps at different simulation stages. As before, a smaller fidelity parameter $t_r$ (with other fidelity parameters held constant) yields more accurate simulations at higher computational costs, with $t_r = 0$ denoting the highest (limiting) fidelity level. Let $\eta(\mathbf{x},\mathbf{t})$ denote the deterministic code output at inputs $\mathbf{x}$ and fidelity parameters $\mathbf{t}$. The CONFIG model assumes the following decomposition of $\eta(\mathbf{x},\mathbf{t})$:
\begin{equation}
    \eta(\mathbf{x},\mathbf{t}) = \eta(\mathbf{x},\mathbf{0}) + \delta(\mathbf{x},\mathbf{t}) \coloneqq  \phi(\mathbf{x})+\delta(\mathbf{x},\mathbf{t}).
    \label{eq:IO}
\end{equation}
Similar to before, $\phi(\mathbf{x}) := \eta(\mathbf{x},\mathbf{0})$ models the ``exact'' simulation solution at the highest (limiting) fidelity setting of $\mathbf{t} \rightarrow \mathbf{0}$, and $\delta(\mathbf{x},\mathbf{t})$ models the numerical discrepancy (or error) between the exact solution $\phi(\mathbf{x})$ and the simulated output $\eta(\mathbf{x},\mathbf{t})$. Since both $\phi(\mathbf{x})$ and $\delta(\mathbf{x},\mathbf{t})$ are unknown, we again place independent Gaussian process priors on both terms. For $\phi(\mathbf{x})$, a standard GP is assigned with user-defined basis functions for the mean and a stationary correlation function. In a later implementation, we will make use of linear basis functions along with the popular squared-exponential correlation function:
\begin{equation}
    \text{Cov}[\phi(\mathbf{x}_1),\phi(\mathbf{x}_2)]=\sigma_1^2K_{\mathbf{x}}^\phi(\mathbf{x}_1,\mathbf{x}_2)=\sigma_1^2\text{exp}\left\{-\sum_{s=1}^p \gamma_s (x_{1,s}-x_{2,s})^2 \right\},
    \label{eq:cov_phi}
\end{equation}
where $\gamma_s$ is the weight parameter for the $s^{\text{th}}$ input dimension.

For the bias term $\delta(\mathbf{x},\mathbf{t})$, we will carefully specify a new non-stationary covariance function that captures one's \textit{prior} knowledge on the numerical convergence behavior. One desirable property of $\delta(\mathbf{x},\mathbf{t})$ is the limiting constraint:
\begin{equation}
    \lim_{\mathbf{t}\rightarrow\mathbf{0}} \; \delta(\mathbf{x},\mathbf{t})=0, \quad \text{for all $\mathbf{x} \in [0,1]^p$ almost surely.}
    \label{eq:limiting}
\end{equation}
In words, for any inputs $\mathbf{x}$, the simulation output $\eta(\mathbf{x},\mathbf{t})$ should converge to the underlying exact solution $\phi(\mathbf{x})$ when \textit{all} fidelity parameters converge to zero, i.e., all fidelity levels are set to their highest (limiting) setting. Property \eqref{eq:limiting} should thus be satisfied almost surely if the simulator enjoys theoretical convergence guarantees (e.g., weak convergence of PDE solutions) or is trusted to converge empirically. Another desirable property is that, for a fidelity parameter $t_r$ and fixed levels of the remaining fidelity parameters $\mathbf{t}_{-r} \neq \mathbf{0}$, we have:
\begin{equation}
    \lim_{t_r\rightarrow 0} \; \delta(\mathbf{x},\mathbf{t}) \neq 0, \quad \text{for all $\mathbf{x} \in [0,1]^p$ almost surely.}
    \label{eq:limiting2}
\end{equation}
In words, for any inputs $\mathbf{x}$ and any positive fidelity parameters $\mathbf{t}_{-r}$, there should be a non-negligible discrepancy between the simulation output $\eta(\mathbf{x},\mathbf{t})$ and the underlying true solution $\phi(\mathbf{x})$. This is again intuitive when the variables in $\mathbf{t}_{-r}$ are fidelity parameters since the simulator should not be expected to reach the true solution when some of these parameters are not at their highest fidelities, i.e., $\mathbf{t}_{-r} \neq \mathbf{0}$. The two limiting constraints thus describe how fidelity parameters determine the discrepancy behavior of the simulator: only when \textit{all} fidelity parameters approach zero should the simulator converge to the true solution.

To satisfy these two properties, we place a Gaussian process prior on $\delta(\mathbf{x},\mathbf{t})$ with product covariance form:
\begin{equation}
    \text{Cov}[\delta(\mathbf{x_1},\mathbf{t}_1),\delta(\mathbf{x_2},\mathbf{t}_2)]= \sigma_2^2 K_{\mathbf{x}}^\delta(\mathbf{x}_1,\mathbf{x}_2)K_{\mathbf{t}}(\mathbf{t}_1,\mathbf{t}_2),
    \label{eq:cov_delta}
\end{equation}
i.e., the effect of input variables and fidelity parameters are assumed to be separable for $\delta$. For the first kernel $K_{\mathbf{x}}^\delta(\cdot,\cdot)$, one can employ a standard stationary kernel; we make use of the squared-exponential kernel:
\begin{equation}
    K_{\mathbf{x}}^\delta(\mathbf{x}_1,\mathbf{x}_2) = \text{exp}\left\{-\sum_{s=1}^p \alpha_s (x_{1,s}-x_{2,s})^2 \right\}
\end{equation}
in our later implementation. For the second kernel $K_{\mathbf{t}}(\mathbf{t}_1,\mathbf{t}_2)$, a careful \textit{non-stationary} specification is needed to satisfy the aforementioned two properties; one can show that this non-stationarity is necessary but not sufficient for satisfying these properties, see \cite{tuo2014surrogate} and later discussion. We will present the next two choices for this kernel, which cater to the two common scenarios for conglomerate multi-fidelity simulators from Section \ref{sec:cong}.


We note that these kernel choices are but recommendations -- the modeler should carefully consider prior domain knowledge to carefully select a kernel that captures such knowledge. With the kernel $K_{\mathbf{t}}$ specified (along with kernels $K_{\mathbf{x}}^\delta$ and $K_{\mathbf{x}}^\phi$), one can show that the response surface $\eta(\mathbf{x},\mathbf{t})$ follows a Gaussian process model, with covariance function:
\small
\begin{equation}
    K_\eta\{(\mathbf{x}_1,\mathbf{t}_1),(\mathbf{x}_2,\mathbf{t}_2)\}\coloneqq \text{Cov}[\eta(\mathbf{x_1},\mathbf{t}_1),\eta(\mathbf{x_2},\mathbf{t}_2)] =\sigma^2_1 K_{\mathbf{x}}^\phi(\mathbf{x}_1,\mathbf{x}_2)+\sigma^2_2 K_{\mathbf{x}}^\delta(\mathbf{x}_1,\mathbf{x}_2)K_{\mathbf{t}}(\mathbf{t}_1,\mathbf{t}_2).
    \label{eq:m2gp_k1}
\end{equation}
\normalsize
The predictive equations for the CONFIG model then follow immediately from the standard GP equations \eqref{eq:gppred} and \eqref{eq:predeqn} with kernel $K_\eta$ given above, with the desired prediction point $(\mathbf{x}^*,\mathbf{0})$ as the goal is to predict the (limiting) highest-fidelity setting. We provide further details on these predictive equations in Section \ref{sec:ml}.




\subsection{Kernel Option 1}
\label{sec:kernel1}
Consider the first kernel choice for $K_\mathbf{t}$ (Kernel 1), which we recommend for Scenario 1 above. This takes the non-stationary form:
\small

\begin{equation}
    K_\mathbf{t}(\mathbf{t}_1,\mathbf{t}_2)= \text{exp}\left\{ -\sum_{r=1}^q \theta_r (t_{1,r}-t_{2,r})^2 \right\}
    -\text{exp}\left\{ -\sum_{r=1}^q \theta_r t_{1,r}^2 \right\}-\text{exp}\left\{ -\sum_{r=1}^q \theta_r t_{2,r}^2 \right\}+1.
    \label{eq:kernel_1}
\end{equation}
\normalsize Here, $\theta_r$ denotes the weight parameter for the $r^{\text{th}}$ fidelity parameter. A larger $\theta_r$ indicates greater sensitivity of discrepancy $\delta$ to the $r^{\text{th}}$ fidelity parameter, and vice versa. One can check that, with this kernel \eqref{eq:kernel_1}, the two desired properties \eqref{eq:limiting} and \eqref{eq:limiting2} for $\delta$ are satisfied, meaning such a kernel indeed captures the aforementioned prior information on numerical convergence behavior.








Kernel 1 is inspired by the non-stationary covariance function in \cite{gul2018uncertainty}, which was proposed for a different task of uncertainty propagation for system outputs. The rationale for this kernel here is as follows. For simplicity, let $\delta(\mathbf{t})$ denote the bias term at some fixed input $\mathbf{x}$. One way to ensure $\delta(\mathbf{t})$ satisfies the limiting condition \eqref{eq:limiting}, i.e., $\lim_{\mathbf{t} \rightarrow \mathbf{0}}\delta(\mathbf{t}) = 0$, is to represent it as a difference of two terms:
\begin{equation}
\delta(\mathbf{t}) = \kappa(\mathbf{t}) - \kappa(\mathbf{0}),
\label{eq:gul1}
\end{equation}
where $\kappa(\cdot)$ can be modeled as a GP. In words, the limiting condition on $\delta(\cdot)$ is enforced by centering $\kappa$ by its response at the limiting fidelity $\mathbf{0}$. The covariance function for $\delta$ can then be written as:
\begin{align}
\small
\begin{split}
\text{Cov}[\delta(\mathbf{t}_1), \delta(\mathbf{t}_2)] &= \text{Cov}[\kappa(\mathbf{t}_1)-\kappa(\mathbf{0}),\kappa(\mathbf{t}_2)-\kappa(\mathbf{0})]\\
&= \text{Cov}[\kappa(\mathbf{t}_1),\kappa(\mathbf{t}_2)] - \text{Cov}[\kappa(\mathbf{t}_1),\kappa(\mathbf{0})] - \text{Cov}[\kappa(\mathbf{t}_2),\kappa(\mathbf{0})] + \text{Cov}[\kappa(\mathbf{0}),\kappa(\mathbf{0})].
\label{eq:gul3}
\end{split}
\end{align}
Kernel 1 in \eqref{eq:kernel_1} can be recovered from \eqref{eq:gul3} with a squared-exponential correlation function on $\kappa$, and satisfies the desired limiting condition \eqref{eq:limiting} by construction.


Kernel 1 has several appealing features for conglomerate multi-fidelity emulation. First, in many applications, one may have prior knowledge of the \textit{continuity} of the underlying numerical solutions (e.g., from FEA theory). With Kernel 1, the corresponding prior process on discrepancy $\delta$ can be shown to yield continuous sample paths, thus capturing such prior knowledge from a Bayesian perspective. Second, the form of this kernel provides a flexible framework for modeling \textit{interactions} between fidelity parameters across different stages. Compared to the additive structure in Kernel 2 introduced later, the latent GP model on $\kappa(\cdot)$ (with the squared-exponential kernel) provides a flexible framework for learning interactions between different fidelity parameters. Because of this, Kernel 1 appears to work best in Scenario 1 for emulating a \textit{single} mechanism with multiple fidelity parameters, e.g., the FEA for beam deflection with different fidelities for each dimension, as such systems often have significant interaction effects between fidelity parameters, e.g., between mesh sizes of each dimension.

\subsection{Kernel Option 2}
\label{sec:kernel2}

Consider next the second choice for $K_\mathbf{t}$ (Kernel 2), which we recommend for the multi-stage \textit{sequential} simulations in Scenario 2. This kernel takes the non-stationary form:

\begin{equation}
    K_\mathbf{t}(\mathbf{t}_1,\mathbf{t}_2) = \left[\sum_{r=1}^q \theta_r\text{min}(t_{1,r},t_{2,r})^{l_r}\right]^l.
    \label{eq:kernel_2}
\end{equation}
Here, $\theta_r$ is a weight parameter for the $r^{\text{th}}$ fidelity parameter, and $l_r$ and $l$ are kernel hyperparameters which we discuss later. Similar to Kernel 1, a greater $\theta_r$ allows for greater sensitivity of the discrepancy $\delta$ to the $r^{\text{th}}$ fidelity parameter. We can again show that with this kernel \eqref{eq:kernel_2}, the two properties \eqref{eq:limiting} and \eqref{eq:limiting2} for bias $\delta$ are satisfied. This follows from the observations that $K_{\mathbf{t}}(\mathbf{t}',\mathbf{t}')$ approaches 0 as $\mathbf{t}' \rightarrow \mathbf{0}$, and that given non-zero entries in $\mathbf{t}'$, $K_\mathbf{t}(\mathbf{t}',\mathbf{t}') \neq 0$ (see Appendix for further discussion). Such a kernel choice thus captures the desired prior information on numerical convergence. With Kernel 2, the resulting prior process on discrepancy $\delta$ can be viewed as a multivariate extension of a standard Brownian motion model \citep{durrett2019probability}, and extends the non-stationary model \eqref{eq:nonstationary} in \cite{tuo2014surrogate}, which tackled only the case of one fidelity parameter.


Kernel 2 has several appealing features for conglomerate multi-fidelity emulation, particularly when the multiple stages are performed \textit{sequentially} over time (see Scenario 2 at the start of the section). One can show that the parametrization of this kernel is directly inspired by (and thus can capture prior information on) standard numerical convergence results for multi-stage simulators. To see why, consider first the simple setting of a \textit{single} fidelity parameter $t$, and let $v_0$ and $v_t$ be the exact and simulated solutions at fidelity $t$ respectively. In the case of finite element analysis (where $t$ is the mesh grid size), it is well-known \citep{brenner2008mathematical} that the numerical error of the simulator can be upper bounded as: 
\begin{equation}
    \|\nu_0-\nu_t\| \leq Ct^\xi,
    \label{eqn:error_1d}
\end{equation}
where $||\cdot||$ is an appropriate norm on the solution space, $\xi$ is a rate parameter, and $C$ is a constant. In words, the numerical error resulting from mesh discretization decays polynomially as mesh size $t$ decreases. Similar polynomial decay rates have also been shown for a broad range of fidelity parameters in numerical solvers, e.g., for elliptical PDEs \citep{hundsdorfer2003numerical} and large-eddy simulations in fluid mechanics \citep{templeton2015calibration}. 

Consider now the \textit{multi-stage} simulators from Scenario 2, where a separate phenomenon is simulated sequentially at each stage. Suppose, at stage $r$, its precision is controlled by a fidelity parameter $t_r$. For this parameter $t_r$, further suppose the simulation error at this stage can be bounded by \eqref{eqn:error_1d} with rate parameter $\xi_r$. One example of this is multi-stage finite element simulators when each stage involves a distinct finite element model (FEM) whose precision depends on a mesh size parameter $t_r$. Similar to before, let $\nu_{\mathbf{0}}$ and $\nu_{t_1, \cdots, t_q}$ denote the exact solution and the simulated solution at fidelity parameters $t_1, \cdots, t_q$. Applying the triangle inequality iteratively, the error between $\nu_{t_1, \cdots, t_q}$ and $\nu_{\mathbf{0}}$ can then be bounded as:

\begin{align}
\begin{split}
    ||v_{\mathbf{0}}-v_{t_1,\cdots,t_q}|| & \leq ||v_{\mathbf{0}}-v_{t_1,0, \cdots, 0}|| + ||v_{t_1,0, \cdots, 0}-v_{t_1,t_2, 0, \cdots, 0}|| + \cdots\\ &+ ||v_{t_1,\cdots,t_{q-1},0}-v_{t_1,\cdots,t_{q-1},t_q}||\\
    &\leq \sum_{r=1}^q C_rt_r^{\xi_r},
    \label{eqn:error_md}
\end{split}
\end{align}
where $C_1, \cdots, C_q$ are again constants. We now show that Kernel 2 indeed captures the error bound \eqref{eqn:error_md} as \textit{prior information} within its kernel specification. To see why, consider the prior standard deviation of the discrepancy term $\delta(\mathbf{x},\mathbf{t})$. From a Bayesian modeling perspective, this should capture the modeler's prior belief on the expected numerical error of the simulator. With $K_{\mathbf{t}}$ set as Kernel 2, one can show that this prior standard deviation takes the form:

\begin{equation}
    \sqrt{\text{Var}\left\{\delta(\mathbf{x},\mathbf{t})\right\}} =\sigma_2 \left[\sum_{r=1}^q \theta_r t_r^{l_r}\right]^{l/2}.
    \label{eqn:priorsd}
\end{equation}
Comparing \eqref{eqn:priorsd} with \eqref{eqn:error_md}, we see that they are precisely the same with the kernel hyperparameters set as $l=2$ and $l_r = \xi_r$ for $r = 1, \cdots, q$. This suggests that with $K_{\mathbf{t}}$ chosen as Kernel 2, the resulting prior model on discrepancy $\delta(\mathbf{x},\mathbf{t})$ indeed captures (on expectation) the numerical error convergence of the multi-stage simulator.

The above connection also helps guide how the specification of hyperparameters for Kernel 2. If the rate parameters $\xi_1, \cdots, \xi_q$ can be identified via a careful analysis of the error bound \eqref{eqn:error_1d} at each stage, one can use simply set the hyperparameters as $l_r = \xi_r$ for $r = 1, \cdots, q$. However, for more complex multi-stage simulators, one may not be able to identify the precise error convergence rates at each stage. In such cases, the kernel hyperparameters can be estimated via maximum likelihood or a fully Bayesian approach (see Section \ref{sec:paraminf}) or set at a fixed value (e.g., $l_r=l=2$). Whether such hyperparameters are set a priori or inferred from data, the infusion of such prior information can yield noticeably improved predictive performance for multi-fidelity emulation, as we show later in Section \ref{sec:multistage}.

It is worth noting that, with the Brownian-like Kernel 2, sample paths from the discrepancy process $\delta(\mathbf{x},\mathbf{t})$ will be highly non-smooth. In particular, within any neighborhood around $\mathbf{t} = \mathbf{0}$, the discrepancy $\delta(\mathbf{x},\mathbf{t})$ will equal 0 an infinite number of times. This may be unintuitive for other properties of discretization error (see, e.g., Equation (1) of \cite{bect2021quantification}), which require that $\delta(\mathbf{x},\mathbf{t}) = 0$ only when $\mathbf{t} = 0$. Our justification for Kernel 2 is not from such properties, but rather from its ability to embed prior information on \textit{expected} numerical convergence via its non-stationary specification. In applications where trajectory smoothness is a concern, Kernel 1 may be a better kernel choice; more on this below.

\begin{figure*}[!t]
     \centering
     \begin{subfigure}[b]{0.45\textwidth}
         \centering
         \includegraphics[width=\textwidth]{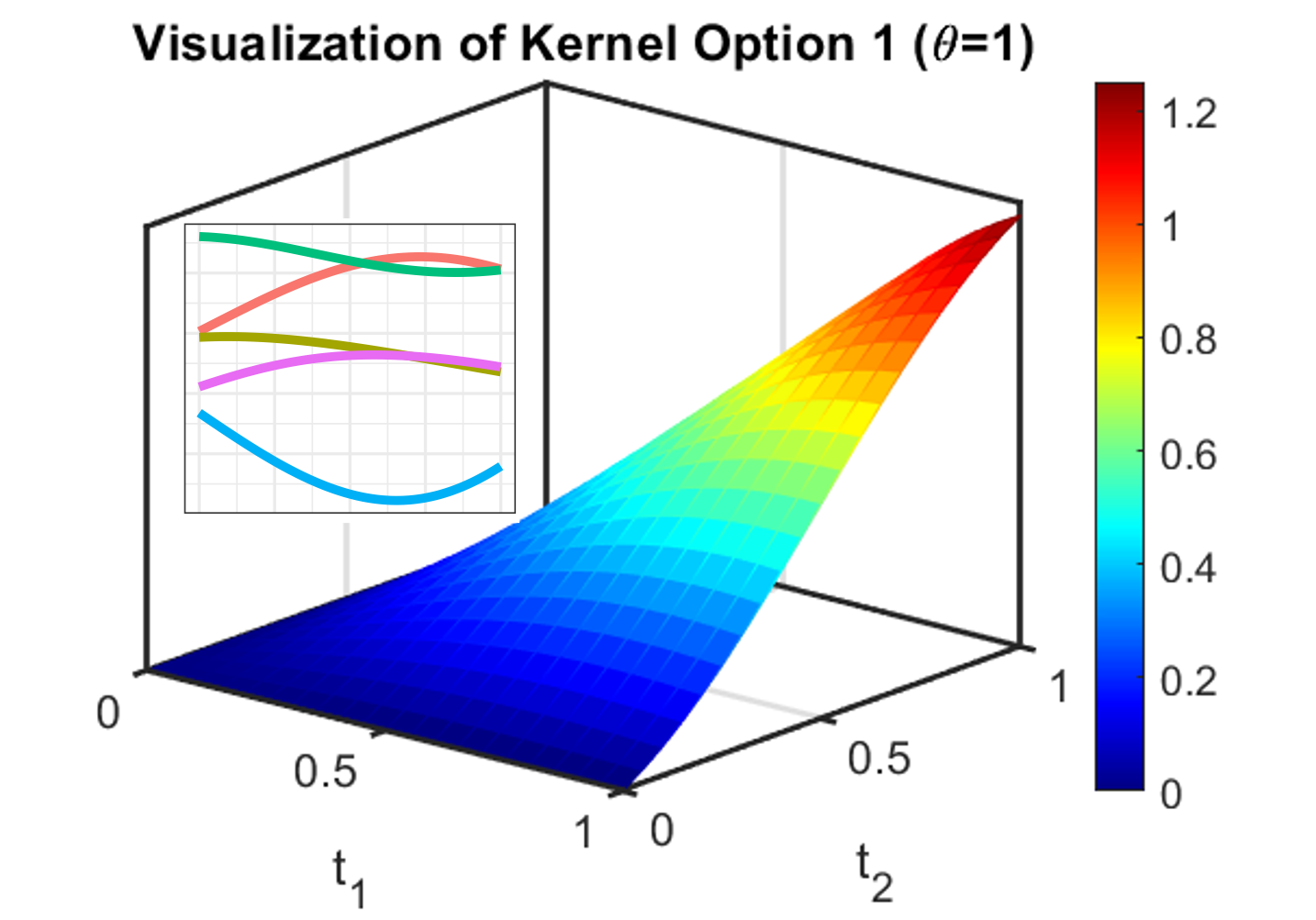}
         \caption{Visualizing Kernel 1 \eqref{eq:kernel_1} with corresponding sample paths using $q=1$ and $\theta=1$.}
         \label{fig:Kernel_vis_1}
     \end{subfigure}
     \hfill
     \begin{subfigure}[b]{0.45\textwidth}
         \centering
         \includegraphics[width=\textwidth]{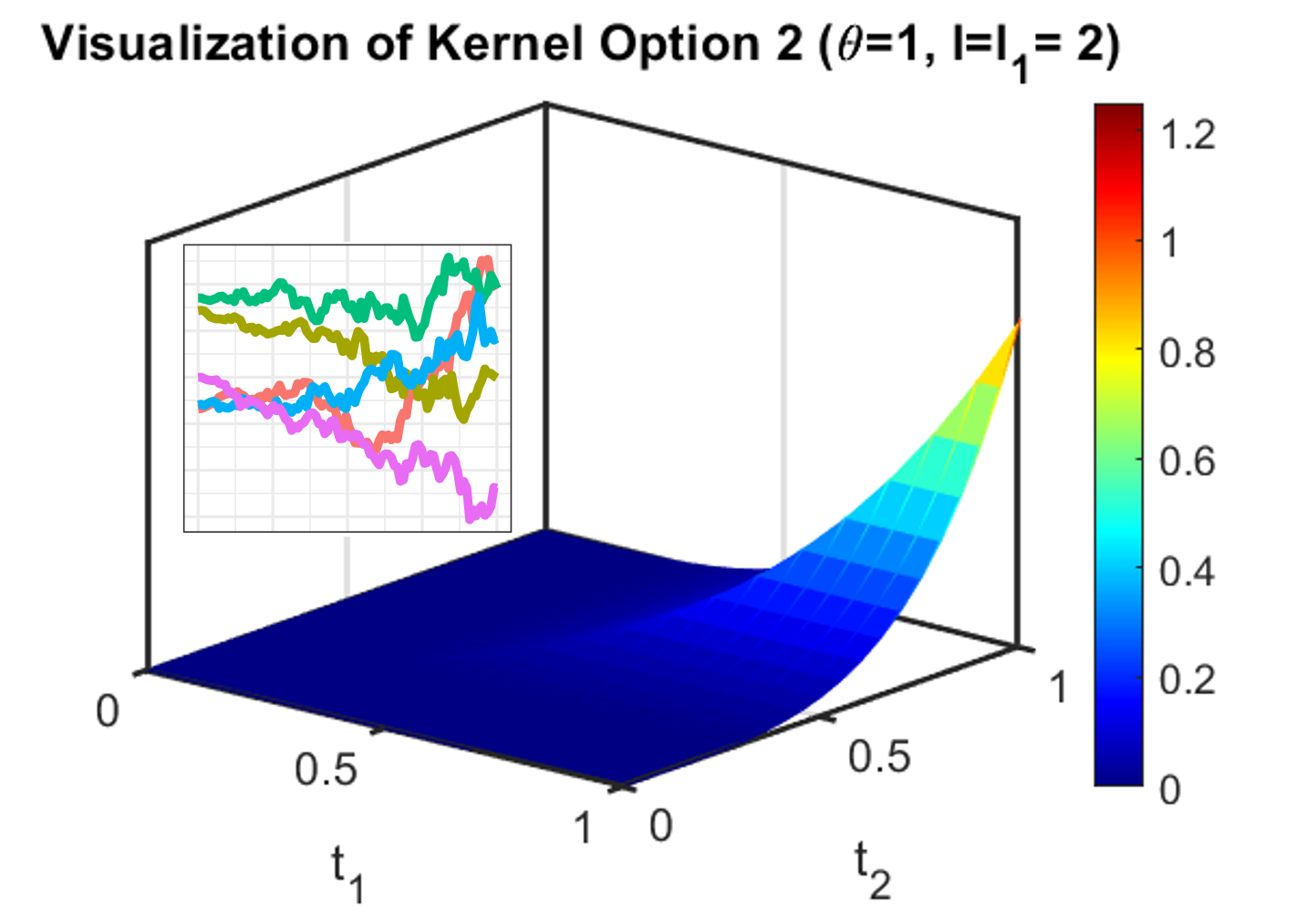}
         \caption{Visualizing Kernel 2 \eqref{eq:kernel_2} with corresponding sample paths} using $q=1$, $\theta=1$, $l=l_1=2$.
         \label{fig:Kernel_vis_2}
     \end{subfigure}
    \caption{Visualization of both kernel options (with corresponding sample paths) for CONFIG with a single fidelity parameter $t$.}
    \label{fig:Kernel_vis}
\end{figure*}

Figure~\ref{fig:Kernel_vis} visualizes the two proposed nonstationary kernels in the simple setting with a single fidelity parameter $t$. For Kernel 1, we set $\theta_q = 1$ and $q=1$, and for Kernel 2, we set $l=l_q=2$ and $q=1$. We see these two kernels have noticeably different shapes: Kernel 1 shows a smooth and gradual increase as either $t_1$ or $t_2$ increases, whereas Kernel 2 exhibits a sharper increase and has a cusp along the line $t_1=t_2$. This cusp causes the highly non-smooth sample paths from Kernel 2, whereas the smoother Kernel 1 induces smoother sample paths, as can be seen from the corresponding sample paths in Figure \ref{fig:Kernel_vis}.

\subsection{Kernel Recommendation}

We provide next a concise summary of kernel recommendation for the CONFIG model. In applications where the conglomerate simulator uses multiple fidelity parameters (e.g., spatial mesh size or temporal discretization) for simulating a single mechanism (Scenario 1), we recommend the use of Kernel 1 \eqref{eq:kernel_1}, which can better account for stronger interactions between different fidelity parameters. We will encounter such an application in Section \ref{sec:beam}. On the other hand, in applications where the conglomerate simulator comprises of multiple sequential stages that model for separate mechanisms, we recommend the use of Kernel 2 \eqref{eq:kernel_2}, which can be justified via numerical error analysis for these simulators. Our motivating QGP application falls within this setting, which we will investigate further in Section \ref{sec:multistage}.

There are, of course, applications that may not cleanly fall within the two presented scenarios; in such cases, careful consideration is needed for an informed kernel specification. In our later numerical experiments, we have found that when there is little prior knowledge on the degree of interaction between fidelity parameters, Kernel 2 seems to be a much more robust choice for predictive modeling; we would thus recommend Kernel 2 for such problems.



\section{Implementation}


\label{sec:imp}
We now discuss important implementation details for the CONFIG model. We present two parameter inference approaches, the first via maximum likelihood and the second via a fully Bayesian formulation for incorporating external knowledge and richer uncertainty quantification. We then outline plausible experimental design strategies.

\subsection{Parameter Inference}
\label{sec:paraminf}

\subsubsection{Maximum Likelihood} \label{sec:ml}


We first present a maximum likelihood approach for estimating the CONFIG model parameters. Let $\boldsymbol{\Theta}_{\rm MLE} = (\boldsymbol\beta, \boldsymbol\gamma, \boldsymbol\alpha, \boldsymbol\theta, \sigma_1^2, \sigma_2^2)$ be the set of parameters to infer, where $\boldsymbol\gamma$ is the vector of weight parameters for $K_{\mathbf{x}}^\phi$, $\boldsymbol\alpha$ is the vector of weight parameters for $K_{\mathbf{x}}^\delta$, and $\boldsymbol\theta$ is the vector of weight parameters for the CONFIG kernel $K_{\mathbf{t}}$ (either Kernel 1 or Kernel 2). Here, we presume the hyperparameters $l_r$ in Kernel 2 \eqref{eq:kernel_2} are pre-specified (similar to \cite{tuo2014surrogate}) and thus not included in the parameter set $\boldsymbol{\Theta}_{\rm MLE}$; in the setting where $l_r$ needs to be estimated, we can simply include them in $\boldsymbol{\Theta}_{\rm MLE}$. Since $\eta(\mathbf{x},\mathbf{t})$ can be expressed as a GP with kernel given in \eqref{eq:m2gp_k1}, one can easily obtain an analytic expression for the likelihood function to optimize. More specifically, let the simulated multi-fidelity training data be $\mathbf{y} = (\eta(\mathbf{x}_i,\mathbf{t}_i))_{i=1}^n$, and let the matrix of basis functions for the GP mean be $\mathbf{F}=(\mathbf{f}(\mathbf{x}_1,\mathbf{t}_1)^T;\mathbf{f}(\mathbf{x}_2,\mathbf{t}_2)^T;\cdots;\mathbf{f}(\mathbf{x}_n,\mathbf{t}_n)^T)$, with corresponding coefficients $\boldsymbol{\beta}$. We thus aim to maximize the log-likelihood of the CONFIG model, given by:
\begin{equation}
    \max_{\boldsymbol{\Theta}_{\rm MLE}} \left\{-\frac{1}{2} \log \det \boldsymbol\Sigma  - \frac{1}{2} (\mathbf{y} - \mathbf{F}\boldsymbol\beta)^T \boldsymbol\Sigma^{-1}  (\mathbf{y} - \mathbf{F}\boldsymbol\beta)\right\},
    \label{eq:log-likelihood}
\end{equation}
where $\det \boldsymbol\Sigma $ is the determinant of the covariance matrix $\boldsymbol\Sigma$.

While the optimization problem \eqref{eq:log-likelihood} is quite high-dimensional, standard non-linear optimization algorithms, such as the L-BFGS-B method \citep{nocedal1999numerical} appears to work well. One can further speed up this optimization procedure via an informed initialization of the parameters $\boldsymbol\Theta_{\rm MLE}$. In particular, we have found that the correlation parameters $\boldsymbol{\alpha}$ can be well-initialized by first fitting a standard Gaussian process model with kernel $K_{\mathbf{x}}^\delta$ over the full training data (ignoring fidelity parameters). With these initial estimates, we then perform the L-BFGS-B non-linear optimization, as implemented in the \textsf{R} package \texttt{stats} \citep{byrd1995limited}.


After maximum likelihood estimation, we would ideally like to integrate such estimates along with their uncertainties within the GP predictive equations \eqref{eq:gppred}, to predict the limiting highest-fidelity surface $\eta(\mathbf{x}^*,\mathbf{0})$ at a new input $\mathbf{x}^*$. However, this integration of uncertainty is difficult to do in closed form for all parameters (see \citep{santner2003design}). We can, however, integrate estimation uncertainty on the mean coefficients $\boldsymbol{\beta}$ in an efficient manner. Following \citep{kennedy2000predicting, bect2021quantification}, the CONFIG predictive mean and variance of $\eta(\mathbf{x}^*,\mathbf{0})$ with such uncertainty integrated (denoted as $\hat{\mu}_\mathbf{0}(\mathbf{x}^*)$ and $s^2_\mathbf{0}(\mathbf{x}^*)$, respectively) becomes:
\begin{equation}
\begin{split}
    \hat{\mu}_\mathbf{0}(\mathbf{x}^*) &=
    {\mu(\mathbf{x}^*,\mathbf{0})} + \mathbf{k}(\mathbf{x}^*,\mathcal{D})^T\mathbf{K}(\mathcal{D})^{-1}(\mathbf{y}_{\mathcal{D}}-\boldsymbol{\mu}(\mathcal{D})),\\
    s^2_\mathbf{0}(\mathbf{x}^*) &= k(\mathbf{x}^*,\mathbf{x}^*)-\mathbf{k}(\mathbf{x}^*,\mathcal{D})^T\mathbf{K}(\mathcal{D})^{-1}\mathbf{k}(\mathbf{x}^*,\mathcal{D}) + (\mathbf{f}(\mathbf{x}^*,\mathbf{0})-\mathbf{k}(\mathbf{x}^*,\mathcal{D})^T\mathbf{K}(\mathcal{D})^{-1}\mathbf{F})^T\\
    &\quad (\mathbf{F}^T\mathbf{K}(\mathcal{D})^{-1}\mathbf{F})^{-1}(\mathbf{f}(\mathbf{x}^*,\mathbf{0})-\mathbf{k}(\mathbf{x}^*,\mathcal{D})^T\mathbf{K}(\mathcal{D})^{-1}\mathbf{F}).
\end{split}
\label{eq:MLE_UQ}
\end{equation}
Unknown model parameters in \eqref{eq:MLE_UQ} can then be plugged in via the maximum likelihood estimates \eqref{eq:log-likelihood}. With this, we can then construct the 95\% predictive interval on the limiting highest-fidelity output $\eta(\mathbf{x}^*,\mathbf{0})$ as:
\begin{equation}
    \left(\hat{\mu}_\mathbf{0}(\mathbf{x}^*)-1.96 \sqrt{s^2_\mathbf{0}(\mathbf{x}^*)},\ \hat{\mu}_\mathbf{0}(\mathbf{x}^*)+1.96 \sqrt{s^2_\mathbf{0}(\mathbf{x}^*)}\right).
\end{equation}




\subsubsection{Fully Bayesian Inference}
\label{sec:Bayesian}

In situations where a richer quantification of uncertainty is desired, a fully Bayesian approach to parameter inference may be appropriate. Below, we present one such approach for the CONFIG model which leverages a Metropolis-within-Gibbs algorithm \citep{gelman1995bayesian} for posterior sampling. 
For an easier derivation of the full conditional distributions, we consider a reparametrization of the covariance kernel \eqref{eq:m2gp_k1} for $\eta(\mathbf{x},\mathbf{t})$ as:
\begin{equation}
    K_\eta\{(\mathbf{x}_1,\mathbf{t}_1),(\mathbf{x}_2,\mathbf{t}_2)\}=\sigma^2\left\{K_{\mathbf{x}}^\phi(\mathbf{x}_1,\mathbf{x}_2)+\lambda K_{\mathbf{x}}^\delta(\mathbf{x}_1,\mathbf{x}_2)K_{\mathbf{t}}(\mathbf{t}_1,\mathbf{t}_2)\right\},
\end{equation}
where $\sigma^2\coloneqq\sigma_1^2$ and $\lambda\coloneqq{\sigma_2^2}/{\sigma_1^2}$.
Here, the new parameter $\lambda$ captures the degree of non-stationarity in the kernel from the influence of the fidelity parameters $\mathbf{t}$. When $\lambda = 0$, the covariance kernel becomes a stationary kernel that depends on only input parameters $\mathbf{x}$.



With this reparametrization, the parameter set to infer is given by $\boldsymbol{\Theta}_{\rm B}=(\boldsymbol\beta,\boldsymbol\gamma,\boldsymbol\alpha,\boldsymbol\theta,\sigma^2,\lambda)$. It is straightforward to show that:
\begin{equation}\label{eq:MCMC_likelihood}
    \mathbf{y}|\boldsymbol{\Theta}_{\rm B} \sim \mathcal{N}(\mathbf{F}\boldsymbol{\beta},\boldsymbol{\Sigma}),
\end{equation}
using the same notation as in \eqref{eq:log-likelihood}. Table \ref{tbl:MCMC_prior} summarizes the priors assigned on parameters $\boldsymbol{\Theta}_{\rm B}$. As before, this specification does not include $l_r$ for Kernel 2, but one can always leverage a reasonable prior distribution on $l_r$ if information is not known on such parameters.
Here, the prior hyperparameters can either be set via prior information or set in a weakly-informative fashion with $a_\lambda=b_\lambda=1$ and $a=b=0.001$ for the remaining hyperparameters.

\begin{table}[!t]
    \centering
    \scalebox{0.97}{
    \begin{tabular}{ccc}
        \hline\hline
        \textbf{Model} & \textbf{Prior Specification}\\
        \hline
        {CONFIG: \quad $\eta(\mathbf{x},\mathbf{t})\sim \mathcal{GP}\{ \boldsymbol{F}\boldsymbol{\beta},K_\eta(\cdot,\cdot)\}$} & {$[\beta_1,\beta_2,\cdots,\beta_m] \overset{i.i.d.}{\sim} 1$}\\
        {Priors: \quad $[\boldsymbol{\Theta}_{\rm B}]=[\boldsymbol\beta
        ][\lambda][\sigma^2|\lambda][\boldsymbol\gamma][\boldsymbol\alpha][\boldsymbol\theta]$} & \\
        \textit{Non-stationary parameter} & {$\lambda \sim \text{Beta}(a_\lambda,b_\lambda)$}\\
        \textit{Kernel precision} & {$1/\sigma^2|\lambda \sim \text{Gamma}(a_\sigma, (1+\lambda)b_\sigma)$}\\
        \textit{Weight parameters} & {$\gamma_1,\gamma_2,\cdots,\gamma_p \overset{i.i.d.}{\sim} \text{Gamma}(a_\gamma, b_\gamma)$}\\
        \textit{Weight parameters} & {$\alpha_1,\alpha_2,\cdots,\alpha_p \overset{i.i.d.}{\sim} \text{Gamma}(a_\alpha, b_\alpha)$}\\
        \textit{CONFIG weight parameters} & {$\begin{cases}
        \theta_1,\theta_2,\cdots,\theta_q \overset{i.i.d.}{\sim} \text{Gamma}(a_\theta, b_\theta) \text{ for Kernel 1}\\
        \theta_1,\theta_2,\cdots,\theta_q \overset{i.i.d.}{\sim} \text{Beta}(a_\theta, b_\theta) \text{ for Kernel 2}
        \end{cases}$}\\
        \hline\hline
    \end{tabular}
    }
    \caption{Hierarchical model specification for the fully Bayesian CONFIG model.}
    \label{tbl:MCMC_prior}
\end{table}

With the priors specified, we now proceed to the posterior sampling algorithm. Of the model parameters in $\boldsymbol{\Theta}_{\rm B}$, we can derive full conditional distributions for two parameters, $\boldsymbol\beta$ and $1/\sigma^2$:
\begin{equation}
    \boldsymbol\beta|\mathbf{y},\boldsymbol\gamma,\boldsymbol\alpha,\boldsymbol\theta,\sigma^2,\lambda \sim \mathcal{N}((\mathbf{F}^T\boldsymbol\Sigma^{-1}\mathbf{F})^{-1}\mathbf{F}^T\boldsymbol\Sigma^{-1}\mathbf{y}, \sigma^2(\mathbf{F}^T\boldsymbol\Sigma^{-1}\mathbf{F})^{-1}),
    \label{eq:MCMC_fullconditionals1}
    \end{equation}
\begin{equation}
    1/\sigma^2|\mathbf{y},\boldsymbol\gamma,\boldsymbol\alpha,\boldsymbol\theta,\lambda,\boldsymbol\beta \sim \text{Gamma}\left(a_\sigma+\frac{n}{2}, (1+\lambda)b_\sigma+\frac{1}{2}(\mathbf{y}-\mathbf{F}\boldsymbol\beta)^T\boldsymbol\Sigma^{-1}(\mathbf{y}-\mathbf{F}\boldsymbol\beta)\right).
        \label{eq:MCMC_fullconditionals2}
\end{equation}
For the remaining parameters in $\boldsymbol{\Theta}_{\rm B}$, we make use of Metropolis-Hastings \citep{metropolis1953equation} steps for sampling the full conditional distributions, as implemented in the \textsf{R} package \texttt{MHadaptive} \citep{chivers2012mhadaptive}. We then iterate these full conditional sampling steps within a Gibbs sampler for posterior exploration of $[\boldsymbol{\Theta}_{\rm B}|\mathbf{y}]$. Algorithm \ref{alg:MHGS} presents the detailed steps for this Metropolis-within-Gibbs sampler for the CONFIG model, with details on burn-in and thinning.


\begin{algorithm}[!t]
\caption{Metropolis-within-Gibbs sampler for the CONFIG model}\label{alg:MHGS}
\hspace*{\algorithmicindent} \textbf{Input:} Training data $\{\mathbf{x}_i,\mathbf{t}_i\}_{i=1}^n$, $\mathbf{y} = (\eta(\mathbf{x}_i,\mathbf{t}_i))_{i=1}^n$; testing input $\mathbf{x}^*$; prior hyperparameters $a_\lambda$, $b_\lambda$, $a_\sigma$, $b_\sigma$, $a_\gamma$, $b_\gamma$, $a_\alpha$, $b_\alpha$, $a_\theta$, $b_\theta$; number of desired MCMC samples $M$; burn-in period $M_{\text{burn-in}} $ and thinning rate $T$.\\
\hspace*{\algorithmicindent} \textbf{Output:} Samples from the posterior distribution $[\boldsymbol{\Theta}_{\rm B} | \mathbf{y}]$.
\begin{algorithmic}[1]
\State Initialize the parameters $\boldsymbol{\Theta}_{\rm B}^{[0]}$ from the prior.
    \For {$\text{iter}=1,\cdots,M_{\text{burn-in}} + TM$}
        \State Sample $\boldsymbol{\beta}^{[\text{iter}]}$ from the full conditional distribution \eqref{eq:MCMC_fullconditionals1}.
        \State Sample $1/\sigma^{2[\text{iter}]}$ from the full conditional distribution \eqref{eq:MCMC_fullconditionals2}.
        \State For the remaining parameters $\{\boldsymbol{\gamma}, \boldsymbol{\alpha},\boldsymbol{\theta},\boldsymbol{\lambda}\}$, perform one step of Metropolis-within-Gibbs sampling using parameters $\{\boldsymbol{\beta}^{[\text{iter}]},1/\sigma^{2[\text{iter}]}\}$.
    \EndFor\\
    Discard the first $M_{\text{burn}}$ samples and thin the remaining samples at a rate of $T$ to obtain samples $\{ \boldsymbol{\Theta}_{\rm B}^{[m]}\}_{m=1}^M$.
\end{algorithmic}
\hspace*{\algorithmicindent} \textbf{Return:} Posterior samples $\{ \boldsymbol{\Theta}_{\rm B}^{[m]}\}_{m=1}^M$ from the posterior distribution $[\boldsymbol{\Theta}_{\rm B} | \mathbf{y}]$.
\end{algorithm}

Finally, with the posterior samples $\{\boldsymbol{\Theta}_{\rm B}^{[m]}\}_{m=1}^M$ obtained from Algorithm \ref{alg:MHGS}, we can easily estimate the posterior predictive mean at a new test point $\mathbf{x}^*$ by marginalizing over $\boldsymbol{\Theta}_{\rm B}$:
\[ \mathbb{E}[\eta(\mathbf{x}^*, \mathbf{0}) | \mathbf{y}] \approx \frac{1}{M}\sum_{m=1}^{M} \hat{\eta}(\mathbf{x}^*,\mathbf{0}|\boldsymbol{\Theta}_{\rm B}^{[m]}), \]
where $\hat{\eta}(\mathbf{x}^*,\mathbf{0}|\boldsymbol{\Theta}_{\rm B}^{[m]})$ is the closed-form GP predictive mean in \eqref{eq:gppred} with fixed hyperparameters $\boldsymbol{\Theta}_{\rm B}^{[m]}$. This serves as the emulator for the fully Bayesian CONFIG model. One can also quantify its uncertainty via posterior predictive samples on $\eta(\mathbf{x}^*, \mathbf{0}) | \mathbf{y}$. These can be obtained by sampling a batch of samples from the predictive distribution $[\eta(\mathbf{x}^*,\mathbf{0})|\mathbf{y},\boldsymbol{\Theta}_{\rm B}^{[m]}]$ in \eqref{eq:gppred} given parameters $\boldsymbol{\Theta}_{\rm B}^{[m]}$, then repeating this procedure on all posterior samples $\{\boldsymbol{\Theta}_{\rm B}^{[m]}\}_{m=1}^M$.




\subsection{Experimental Design}

Of course, given a fixed and limited computational budget, an experimenter would want to maximize the predictive power of the fitted multi-fidelity emulator model. For Gaussian process models, space-filling designs \citep{joseph2016space} -- which aim to uniformly fill up the design space -- are commonly used and have desirable information-theoretic and predictive properties \citep{johnson1990minimax}. Different notions of space-filling designs have been explored in the literature, including maximin designs \citep{johnson1990minimax,morris1995exploratory}, minimax designs \citep{johnson1990minimax,mak2018minimax} and maximum projection (MaxPro) designs \citep{joseph2015maximum}. 

For the CONFIG model, there are several ways in which one can adapt existing space-filling designing methods. One approach is to (i) adopt a space-filling design over the combined design space of both input parameters $\mathbf{x}$ and fidelity parameters $\mathbf{t}$. Such a design ensures that training points are not only well-spaced out over the input space for prediction at untested settings but also well-spaced out over the fidelity space to better learn the effects of individual fidelity parameters. Another approach might be (ii) a crossed array design \citep{wu2011experiments} between input and fidelity space, which are popular designs for robust parameter design. In such a design, one first generates two space-filling designs, one over the input space and the other over the fidelity space, then takes for the final design all combinations of input and output points. Both designs appear to yield good performance: the designs in (i) are used for our numerical experiments and cantilever beam deflection application, and the designs in (ii) are used for the emulation of the QGP evolution. The problem of optimal experimental design for the proposed non-stationary CONFIG model is quite intriguing, and we aim to pursue this in future work.






\section{Numerical Experiments}
\label{sec:numerical}
We now explore the performance of the proposed CONFIG model in a suite of simulation experiments with multiple fidelity parameters. We compare the CONFIG model with several existing emulator models. The first model is a standard GP emulator with a squared-exponential correlation function on both input parameters $\mathbf{x}$ and fidelity parameters $\mathbf{t}$; one then uses the fitted model to predict at $\mathbf{t} = \mathbf{0}$. We call this model simply the ``standard GP'' emulator. The second model is the TWY model \citep{tuo2014surrogate}, which uses a single fidelity parameter. Since there are multiple fidelity parameters in the target conglomerate problem, we will first compute the arithmetic or geometric mean of the fidelity parameters $t_1, \cdots, t_q$, then apply the TWY model with this single aggregate fidelity parameter. We call the resulting models the TWY (ARITH) and TWY (GEOM) emulators, respectively. 

In the following, we investigate the performance of these models on multi-fidelity extensions of two test functions, the 2D Currin function \citep{currin1991bayesian} and the 4D Park function \citep{cox2001statistical}. For CONFIG, we follow Section \ref{sec:imp} and set the power parameters in Kernel 2 as $l_r=l=2$. Kernel hyperparameters for our model are estimated via maximum likelihood in Sections \ref{sec:currin} and \ref{sec:park}, and its fully Bayesian counterpart is explored in Section \ref{sec:bayes_res}.

\subsection{Multi-Fidelity Currin Function}
\label{sec:currin}



\begin{figure*}[!t]
     \centering
     \begin{subfigure}[b]{0.495\textwidth}
         \centering
         \includegraphics[width=\textwidth]{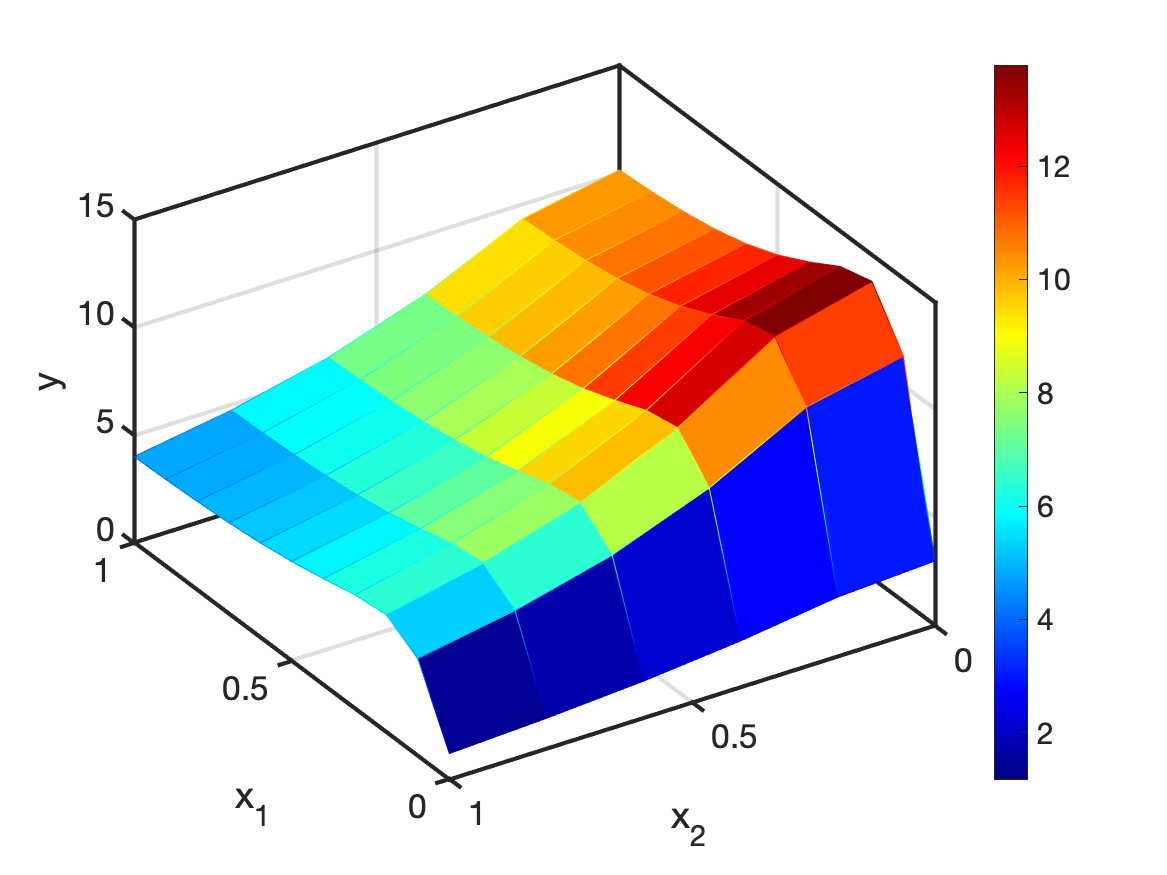}
         \caption{$\eta(\mathbf{x},\mathbf{t})$ with $t_1=0.1$ and $t_2=0.2$.}
         \label{fig:Ex1_vis_1}
     \end{subfigure}
     \hfill
     \begin{subfigure}[b]{0.495\textwidth}
         \centering
         \includegraphics[width=\textwidth]{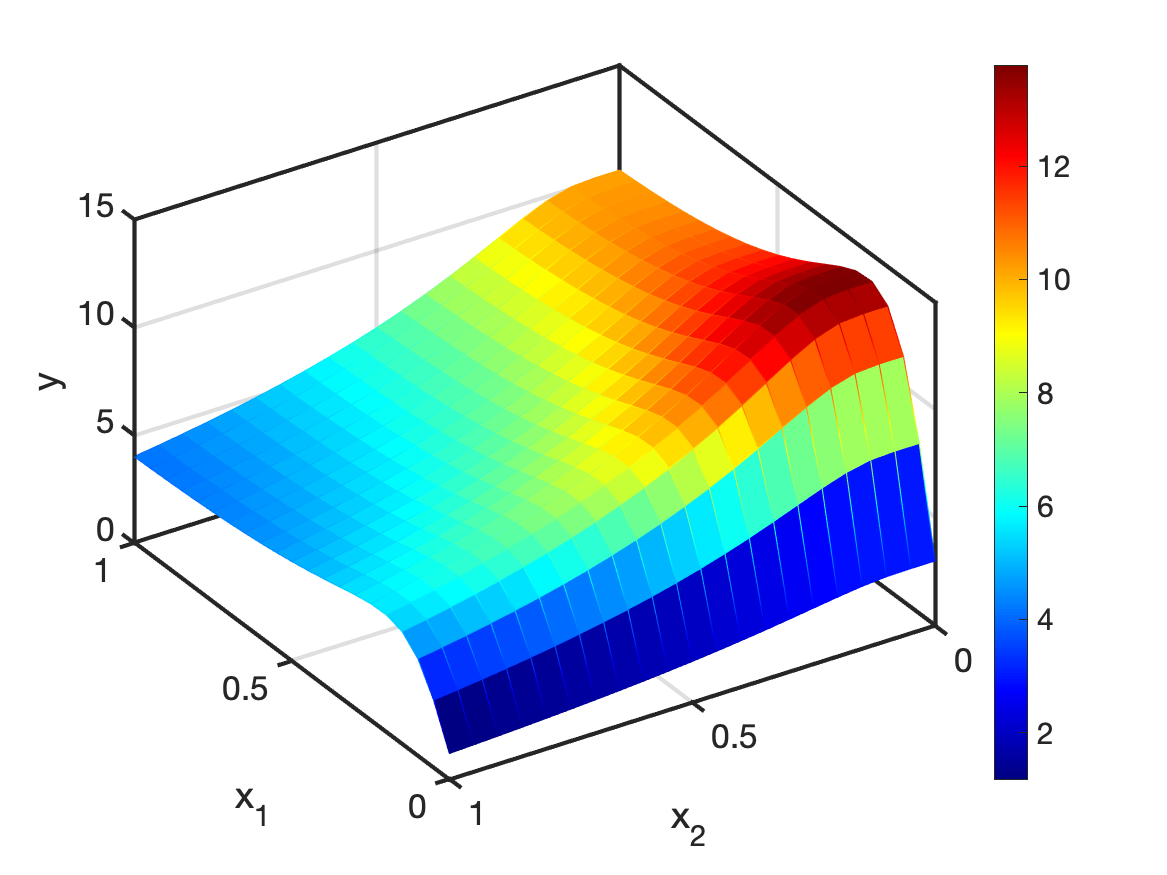}
         \caption{$\eta(\mathbf{x},\mathbf{t})$ with $t_1=0.05$ and $t_2=0.05$.}
         \label{fig:Ex1_vis_2}
     \end{subfigure}
    \caption{Visualization of the multi-fidelity Currin function at two different fidelity settings.}
    \label{fig:Ex1_vis}
\end{figure*}

Our first test function builds off of the 2D Currin test function in \cite{currin1991bayesian}:
\begin{equation}
    \phi(\mathbf{x})=\left[1-\text{exp}\left(-\frac{1}{2x_2}\right)\right]\frac{2300x_1^3+1900x_1^2+2092x_1+60}{100x_1^3+500x_1^2+4x_1+20},
\label{eq:currin}
\end{equation}
where $\mathbf{x}=[x_1,x_2]\in[0,1]^2$. We then build a lower-fidelity representation of this function, denoted as $\eta(\mathbf{x},\mathbf{t})$, with two fidelity parameters $\mathbf{t}=[t_1,t_2]$ via piecewise grid interpolation. More specifically, this approximation is carried out in two steps. First, we generate a rectangular grid in the input space, where the dimension of each mesh cell is $t_1 \times t_2$. Next, we evaluate the underlying function \eqref{eq:currin} at the mesh grid points, and perform piecewise grid interpolation to construct a lower-fidelity version of \eqref{eq:currin}. This procedure is effectively the same as finite element meshing, which splits the input domain into many smaller elements. Figure \ref{fig:Ex1_vis} visualizes this test function $\eta(\mathbf{x},\mathbf{t})$ with $(t_1,t_2)=(0.1,0.2)$ and $(0.05,0.05)$. It is clear that as $t_1$ and $t_2$ become smaller, $\eta(\mathbf{x},\mathbf{t})$ becomes closer to the underlying Currin function \eqref{eq:currin}, which is as desired. 

We then compare the CONFIG emulator with the aforementioned baseline models. For each experiment, we first generate $n=50$ design points over \textit{both} input and fidelity parameters via the MaxPro design \citep{joseph2015maximum}. Here, we set the range for each fidelity parameter to be between 0.1 and 0.4, to mimic the reality that simulations are prohibitively expensive for small choices of fidelity parameters $t_i$. Using this design, we then collect training data from the multi-fidelity Currin function $\eta(\mathbf{x},\mathbf{t})$. For validation, we randomly select $N=\text{1,000}$ points over the input parameter space as the testing set and compare how well these models predict the Currin function $\phi(\mathbf{x})$ in terms of mean squared error (MSE) and its empirical coverage ratio of 95\% predictive intervals. This procedure is then replicated 20 times.



\begin{figure}[!t]
\centering
\begin{minipage}[b]{0.9\textwidth}
        \centering
        \includegraphics[width=1\linewidth]{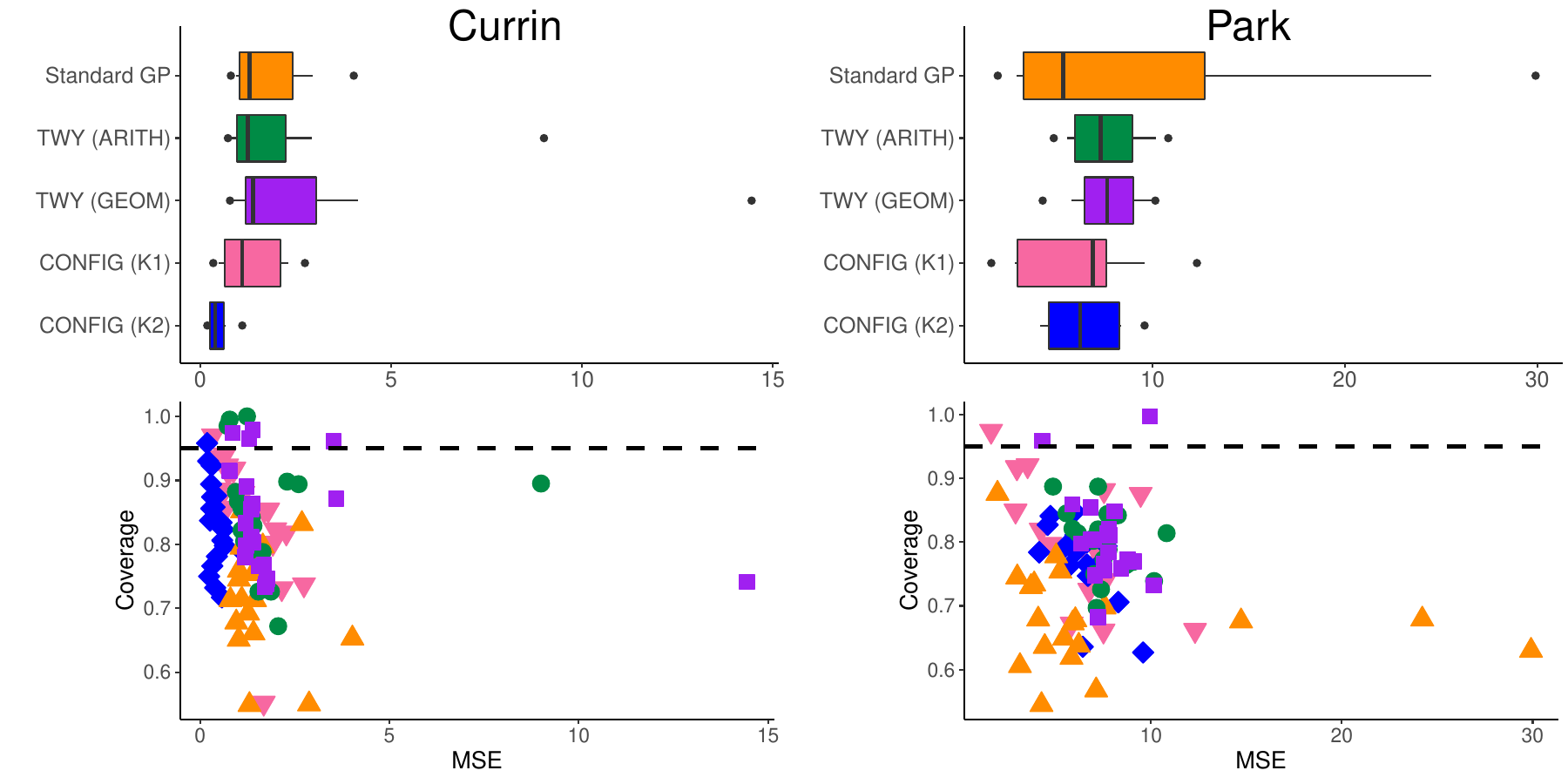}
\end{minipage}
\caption{(Top) Boxplots of MSEs for the multi-fidelity Currin and multi-fidelity Park experiments. (Bottom) Scatterplots of empirical coverage ratios vs. MSEs for the multi-fidelity Currin and multi-fidelity Park experiment. Each dot represents a replication of the experiment, and the black dashed lines denote the nominal 95\% rate.}
\label{fig:currin_park_results}
\vspace{-0.2in}
\end{figure}

\begin{table}[!t]
    \centering
    \begin{tabular}{c|cc|cc}\hline
        \hline
        {Model} & {\makecell{Avg. MSE \\ (Currin)}} & {\makecell{Avg. Coverage \\ (Currin)}} & {\makecell{Avg. MSE \\ (Park)}} & {\makecell{Avg. Coverage \\ (Park)}}\\
        \hline
        {Standard GP} & {1.537} & {71.93\%} & {7.587} & {67.96\%}\\
        {TWY (ARITH)} & {1.772} & {84.46\%} & {7.509} & {79.96\%}\\
        {TWY (GEOM)} & {2.233} & {83.96\%} & {7.609} & {\textbf{80.68\%}}\\
        {CONFIG (Kernel 1)} & {1.310} & {\textbf{84.90\%}} & {\textbf{6.141}} & {80.60\%}\\
        {CONFIG (Kernel 2)} & {\textbf{0.438}} & {83.18\%} & {6.429} & {76.41\%}\\
        \hline\hline
    \end{tabular}
    \caption{Average testing MSEs and empirical coverage ratios for the multi-fidelity Currin and Park experiments over 20 replications.}
    \label{tab:currin_park}
\end{table}



Figure~\ref{fig:currin_park_results} (left) shows the boxplots of testing MSEs and a scatterplot of empirical coverage rates against MSEs for the compared models, with Table \ref{tab:currin_park} (left) reporting its corresponding average MSEs and coverage ratios. 
There are several observations of interest. First, CONFIG (with either Kernel 1 or Kernel 2) outperforms existing models in terms of average MSE. This suggests that, by embedding the underlying conglomerate multi-fidelity structure within the non-stationary kernel specification, the proposed model can indeed provide better emulation over models that do not explicitly integrate this information. The improved performance of CONFIG over the TWY models also suggests that, when \textit{multiple} fidelity parameters are present in the simulator, the use of such information can be useful for reducing emulation error over the TWY models, which aggregate fidelity into a single parameter. Finally, CONFIG with Kernel 2 provides noticeably better performance than Kernel 1; this may be because there is little interaction between the two fidelity parameters under the piecewise grid interpolation of $\eta(\mathbf{x},\mathbf{t})$.

As for coverage ratios (Figure~\ref{fig:currin_park_results} and Table \ref{tab:currin_park}, left), we first note that among the 20 replications, not all empirical coverage ratios can attain the nominal rate of 95\%. This highlights an inherent challenge for multi-fidelity uncertainty quantification, as one requires \textit{extrapolation} beyond the range of simulated fidelity parameters to predict for the (limiting) highest fidelity code. This becomes more pronounced for the considered conglomerate setting with \textit{multiple} fidelity parameters and limited data. We see that the standard GP yields the lowest coverage rates; this is unsurprising, as it fails to account for the non-stationary behavior of fidelity parameters $\mathbf{t}$ from numerical convergence. Both CONFIG and TWY provide comparable coverage rates that are slightly below 95\%, particularly that for CONFIG Kernel 2. We will address this undercoverage later in Section \ref{sec:bayes_res}.

\subsection{Multi-Fidelity Park Function}
\label{sec:park}
Our second test function builds off of the 4D Park test function in  \cite{cox2001statistical}:
\begin{equation}
    \phi(\mathbf{x}) = \frac{x_1}{2}\left[  \sqrt{ 1+(x_2+x_3^2)\frac{x_4}{x_1^2} }-1\right]+(x_1+3x_4)\text{ exp}\left(1+\text{sin}(x_3)\right).
    \label{eq:park}
\end{equation}
We again build a lower-fidelity representation of $\phi(\mathbf{x})$, denoted as $\eta(\mathbf{x},\mathbf{t})$, using four fidelity parameters $\mathbf{t} = (t_1, \cdots, t_4)$ via piecewise grid interpolation. Similar to before, we use MaxPro designs (with $n=50$ design points) over both input and fidelity parameters, with a range of $[0.2,0.5]$ for each fidelity parameter. The same emulator models are compared as before, and the experiment is replicated 20 times over $N = \text{1,000}$ random testing points.

Figure~\ref{fig:currin_park_results} (right) shows the boxplots of testing MSEs and a scatterplot of empirical coverage rates against MSEs for the compared models, with Table \ref{tab:currin_park} (right) reporting its average MSEs and coverage ratios. 
We see again that the proposed CONFIG model outperforms its competitors by a noticeable margin in terms of MSE, which affirms the value of embedding prior information on the conglomerate multi-fidelity simulator within a carefully constructed non-stationary kernel. For coverage ratios, all five models yield lower coverage rates than for the Currin function; this may be due to the increasing challenge of UQ for extrapolation in a higher-dimensional setting. Despite this, CONFIG (with both Kernel 1 and 2) maintains comparable coverage to TWY and higher coverage than the standard GP.

\subsection{Modified Multi-Fidelity Currin Function}
We now explore a more complex modification of the Currin function, which integrates additional structure for the high-fidelity function. Our third test function $\eta_{\text{mod}}(\mathbf{x},\mathbf{t})$ takes the form:
\begin{equation}
    \eta_{\text{mod}}(\mathbf{x},\mathbf{t})=\eta_{\rm currin}(\mathbf{x},\mathbf{t}) + (1-t_1) \,\text{sin}(x_1) + (1-t_2)\, \text{cos}(x_2),
\end{equation}
where $\eta_{\rm currin}(\mathbf{x},\mathbf{t})$ is the multi-fidelity Currin function used in Section \ref{sec:currin}. The two additional terms $\text{sin}(x_1)$ and $\text{cos}(x_2)$ impose structure in the high-fidelity function (i.e., with $t_1=t_2=0$), which gets blurred out at lower fidelities (i.e., as $t_1$ or $t_2$ increase). This reflects scenarios where higher-fidelity refinements of the computer code may reveal additional structure not captured at lower fidelities. The same simulation set-up was used here as in Section \ref{sec:currin}.

Table \ref{tab:mod_currin} shows the average MSEs and coverage ratios over 20 replications. We see again that the CONFIG (with either Kernel 1 or 2) outperforms its competitors in terms of testing MSE. For coverage ratios, we see that while the TWY with arithmetic mean yields the highest coverage ratio, this model also returns the highest predictive error, which is clearly not desirable. The CONFIG with Kernel 1 achieves a comparably high coverage ratio with noticeably less prediction error, and the CONFIG with Kernel 2 yields a much lower prediction error at the cost of slight undercoverage (this can be addressed via the following fully Bayesian implementation).

\begin{table}[!t]
    \centering
    \begin{tabular}{c|cc}\hline
        \hline
        {Model} & {\makecell{Avg. MSE \\ (Modified Currin)}} & {\makecell{Avg. Coverage \\ (Modified Currin)}} \\
        \hline
        {Standard GP} & {2.014} & {64.65\%} \\
        {TWY (ARITH)} & {2.132} & {\textbf{85.48\%}}\\
        {TWY (GEOM)} & {2.073} & {83.02\%}\\
        {CONFIG (Kernel 1)} & {1.559} & {84.36\%}\\
        {CONFIG (Kernel 2)} & {\textbf{0.632}} & {73.12\%}\\
        \hline\hline
    \end{tabular}
    \caption{Average testing MSEs and empirical coverage ratios for the modified multi-fidelity Currin experiment over 20 replications.}
    \label{tab:mod_currin}
\end{table}

\subsection{Fully Bayesian Implementation}
\label{sec:bayes_res}


One reason for the slight undercoverage of CONFIG in previous experiments is it employs plug-in parameters estimated via maximum likelihood (Section \ref{sec:ml}). One solution is to employ a fully Bayesian implementation of the model (Section \ref{sec:Bayesian}); we explore its performance for the earlier Currin and Park experiments. For the MLE approach, its 95\% predictive interval is obtained from the closed-form distribution \eqref{eq:MLE_UQ} with plug-in parameter estimates. For the fully Bayesian approach, we use its 95\% highest-posterior-density interval computed from posterior samples on the predictive distribution $[\eta(\mathbf{x}^*, \mathbf{0}) | \mathbf{y}]$. These samples are obtained via five parallel MCMC chains from Algorithm \ref{alg:MHGS} with random initialization. Each chain was run for 10,000 iterations, with the first 5,000 discarded as burn-in, and the remaining samples thinned by a factor of 50 to reduce autocorrelation. Since the fully Bayesian model is more costly to fit, we demonstrate this on only one set of training/testing data from earlier experiments.



For MCMC, convergence was assessed via the Gelman-Rubin statistic \citep{gelman1992inference}, as implemented in the R package \texttt{coda} \citep{coda}. For the fully Bayesian CONFIG model with Kernel 2, all model parameters have a Gelman-Rubin statistic below 1.2, thus suggesting the MCMC has converged \citep{brooks1998general}. However, for the fully Bayesian CONFIG model with Kernel 1, we encountered very poor mixing performance and numerical instability issues, as the posterior distribution of kernel hyperparameters appears to be highly complex and multi-modal. Since undercoverage seems to be less pronounced for Kernel 1 (see Table \ref{tab:currin_park}), we thus recommend a fully Bayesian implementation for only Kernel 2 to address the aforementioned undercoverage issue with plug-in MLEs.

\begin{table}[!t]
    \centering
    \begin{tabular}{c|cc|cc}\hline
        \hline
        {Model} & {\makecell{MSE \\ (Currin)}} & {\makecell{Coverage \\ (Currin)}} & {\makecell{MSE \\ (Park)}} & {\makecell{Coverage \\ (Park)}}\\
        \hline
        {CONFIG (Kernel 2, MLE)} & {0.621} & {82.30\%} & {6.041} & {78.40\%}\\
        {CONFIG (Kernel 2, Bayesian)} & {0.615} & {86.70\%} & {6.396} & {83.40\%}\\
        \hline\hline
    \end{tabular}
    \caption{Testing MSEs and empirical coverage ratios for the plug-in MLE and fully Bayesian CONFIG model with Kernel 2, in the multi-fidelity Currin and Park experiments with the first set of training/testing data.}
    \label{tab:Bayesian_vs_MLE}
    \vspace{-0.2in}
\end{table}

With this in mind, Table \ref{tab:Bayesian_vs_MLE} summarizes the MSEs and coverage ratios for the plug-in MLE and fully Bayesian CONFIG model using Kernel 2. We see that, while the MSEs of the plug-in MLE approach are quite small, its coverage ratios (82.30\% and 78.40\% for the Currin and Park functions, respectively) are lower than the desired rate of 95\%. This is again unsurprising since plug-in MLEs do not account for parameter estimation uncertainty. The fully Bayesian CONFIG model yields similarly small MSEs but provides noticeably closer coverage to the desired 95\% rate by factoring in posterior uncertainty on parameters. While this still yields slight undercoverage (which is unsurprising since extrapolation with GPs is inherently difficult, particularly with multiple fidelity parameters), we see that a fully Bayesian implementation can indeed provide improved uncertainty quantification for conglomerate multi-fidelity emulation.

\section{Applications}
\label{sec:apply}

Finally, we explore the usefulness of the proposed model in two applications. The first application involves the conglomerate multi-fidelity emulation of a cantilever beam deflecting under stress. The second application is the earlier motivating problem of multi-stage multi-fidelity emulation of the quark-gluon plasma produced in heavy-ion collisions. In both applications, parameter estimation for CONFIG is performed via maximum likelihood (see Section \ref{sec:ml}).

\subsection{Cantilever Beam Deflection}
\label{sec:beam}
\begin{figure}[!t]
    \centering
    \includegraphics[width=.5\textwidth]{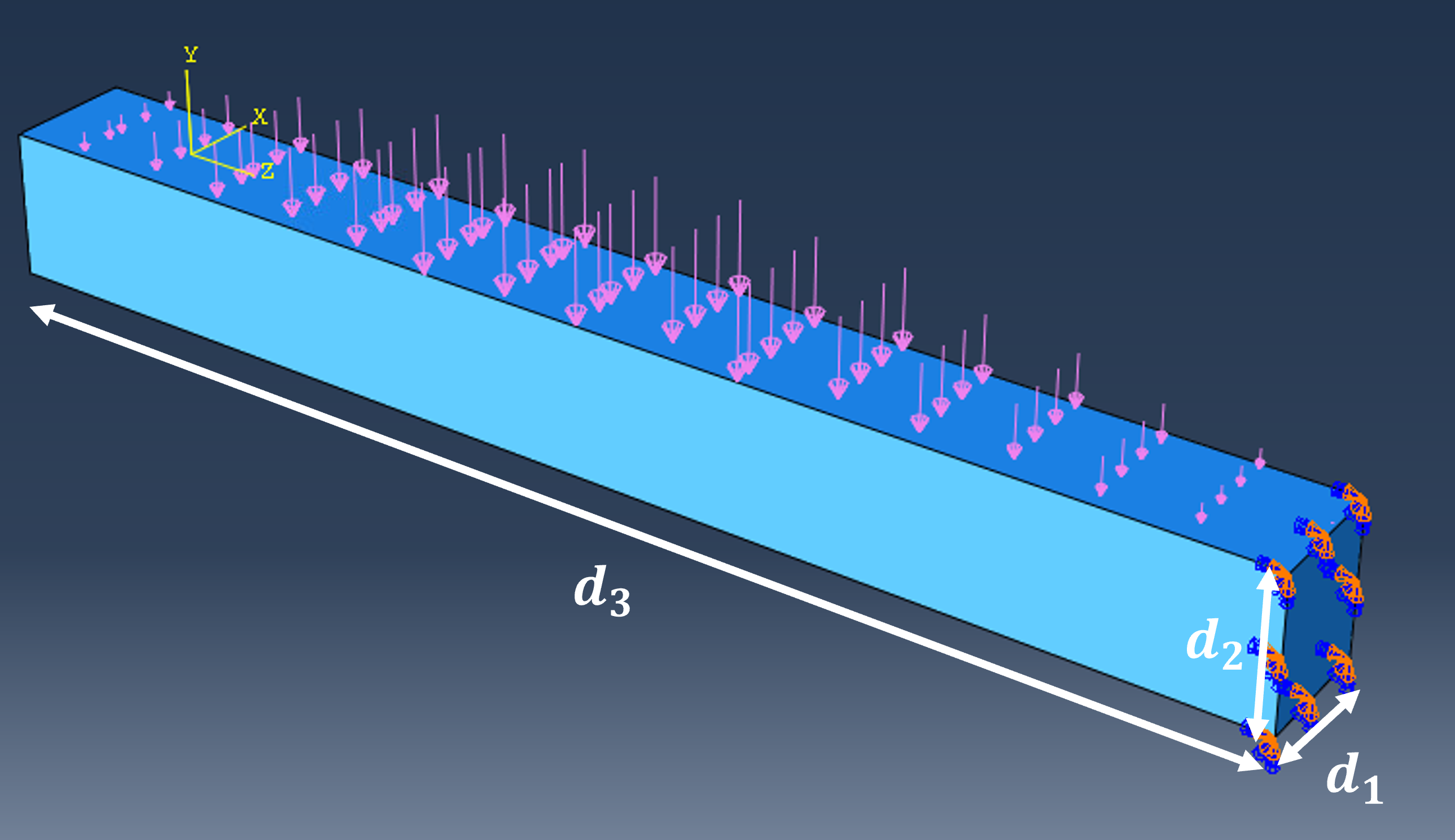}
    \caption{Beam cantilever simulation with fixed end surface as shown in ABAQUS.}
    \label{fig:beam}
\end{figure}

The first application investigates the static stress analysis on a cantilever beam. Beam structures are commonly used in finite elements to model transverse loads and deformation under various circumstances, and the study of their deflection behavior is a canonical problem in finite element analysis and has been studied extensively \citep{ngo1967finite,heyliger1988higher,chakraborty2003new}. Here we use it to evaluate the performance of our modeling framework. Figure~\ref{fig:beam} shows an illustration of the beam for our study, where one end surface of the beam is fixed, and an external pressure field is applied on its top surface. The deflection of this beam under stress is typically simulated using FEA simulations, which can be computationally expensive. For our experiments, these FEA simulations are carried out using the ABAQUS software \citep{smith2009abaqus} with rectangular mesh cells.

The setup is as follows. The beam dimensions are specified by its breadth $d_1$, its height $d_2$, and its length $d_3$. We further let $d_1=d_2$ so the cross-section of the beam is square-shaped (see Figure \ref{fig:beam}). We then set Young's modulus of the beam (which parametrizes the stiffness of the beam) to be 200~MPa and the Poisson ratio (which measures the deformation of the beam under loading) to be 0.28, with material properties corresponding to steel. For the external pressure field, which is applied vertically downward on top of the beam, we employed the continuous half-sine pressure field given by:

\begin{equation}
    p(w) = C_1 x_1 \sin(C_2 w/x_2).
\end{equation}
Here, $w\in[0,d_3]$ denotes the location along the beam from the fixed end, $C_1=2000$ and $C_2=\pi/200$ are constants, $x_1$ is a scale factor for the pressure, and $x_2$ parametrizes the length of the beam, i.e., $d_3=200x_2$. An additional input parameter $x_3$ controls the width and breadth of the beam cross-section $x_3=20d_1=20d_2$. There are thus a total of three input parameters $\mathbf{x}=[x_1,x_2,x_3]\in[0,1]^3$ for this study.

For fidelity parameters, it is natural to consider a meshing procedure that partitions the beam into smaller 3D mesh rectangles. The size of these mesh rectangles can be controlled by three fidelity parameters, which dictate the size of the mesh rectangles in each dimension. In other words, the three fidelity parameters $t_1,t_2,t_3 \in (0,1)$ determine the \textit{scale} of the finite elements. As a result of the pressure field, the cantilever beam will deflect downward, resulting in deflection at its tip. The response of interest is taken to be the amount of tip deflection. The goal is thus to train an emulator model which, using a carefully designed training set of simulation runs over different inputs $\mathbf{x}$ and fidelities $\mathbf{t}$, efficiently predicts the ``exact'' solution for tip deflection (i.e., at $\mathbf{t} = \mathbf{0}$) of a new beam with inputs $\mathbf{x}$.

The experiment is carried out as follows. To generate training data, we first run the simulator (ABAQUS) on a $n=50$-point MaxPro design \citep{joseph2015maximum} over the combined space of input parameters $\mathbf{x}$ and fidelity parameters $\mathbf{t}$, which required about 4.5 hours of computation. For fidelity parameters, we set it to be $\mathbf{t}\in[1/31,1/3]^3$, which ensures we have an integer number of finite elements at the edge case in each dimension (for $\mathbf{t}$ values in between, we round up to the nearest integer). For validation, we further run the simulator on 30 new cantilever configurations (the testing set) where each takes about one hour, uniformly sampled over the input space, to test the performance of each model (in terms of MSE and coverage) in predicting the tip deflections. While the ``exact'' response with $\mathbf{t} = \mathbf{0}$ cannot be obtained numerically, this can be well-approximated by running the simulator at very fine mesh sizes; in our case, we used $\mathbf{t}=[0.025,0.025,0.005]$ for testing points, which provided a sufficiently fine mesh according to a mesh validation study. One simulation run at this high-fidelity setting requires around an hour of computation, meaning there is a considerable opportunity for a multi-fidelity emulator to greatly speed up design exploration.

The same emulators as before (the standard GP, the two TWY models, and the two CONFIG models) are used for comparison. Here, we recommend the use of Kernel 1 for CONFIG, as the application involves the simulation of a single mechanism with multiple fidelity parameters (Scenario 1); Kernel 2 is however included for comparison. We further set $l = l_r = 2$ for CONFIG and $l=2$ for the two TWY models; such a choice captures the fact that the governing deflection equation (between beam deflection and span) involves the derivative of the deflection \cite{tuo2014surrogate}. In addition, we include a ``high-fidelity GP'' emulator model, which is trained on only data from the high-fidelity simulator with $\mathbf{t} = [0.025,0.025,0.005]$. For a fair comparison, this model is trained on high-fidelity points from a four-point MaxPro design, which requires comparable time to simulate as the earlier 50-point designs over the combined input-fidelity space.

\begin{table}[!t]
    \centering
    \begin{tabular}{ccccc}\hline
        \hline
        {Model} & {MSE} & {Average Standard Error} & {Coverage Ratio}\\
        \hline
        {High-fidelity GP} & {475.50} & {28.28} & {\textbf{30/30}}\\
        {Standard GP} & {211.13} & {8.43} & {25/30}\\
        {TWY (ARITH)} & {747.30} & {16.89} & {27/30}\\
        {TWY (GEOM)} & {551.07} & {11.29} & {26/30}\\
        {CONFIG (Kernel 1)} & {\textbf{23.58}} & {\textbf{2.75}} & {26/30}\\
        {CONFIG (Kernel 2)} & {33.18} & {4.77} & {26/30}\\
        \hline\hline
    \end{tabular}
    \caption{Result comparison for the beam deflection application.}
    \label{tab:beam}
\end{table}


Table~\ref{tab:beam} summarizes the MSEs, average standard errors, and empirical coverage ratios (of 95\% predictive intervals) for the compared emulators over 30 test points. Again, we see that CONFIG with the recommended Kernel 1 (but also with Kernel 2) yields noticeably improved predictive performance over existing methods. This again highlights the advantage of an informed kernel specification from the conglomerate multi-fidelity simulator, and is particularly apparent in the cantilever beam application, where the three fidelity parameters bear physical importance. For beam bending, the accuracy of simulations is known to be more sensitive to the mesh density along the beam span (i.e., $d_3$) \citep{cui1992contact}. By explicitly modeling this conglomerate multi-fidelity structure, CONFIG can identify the greater importance of this fidelity parameter via inference of its weight parameter, thus allowing for improved predictions over existing models that ignore such structure.

In terms of coverage ratios, the 95\% predictive intervals for both CONFIG models cover 26 out of 30 test points. Although this is slightly lower than the desired 95\%, this is in line with earlier numerical experiments and may be due to the inherent challenge of multi-dimensional extrapolation with GPs. While the high-fidelity GP and TWY (ARITH) models have higher coverage ratios than CONFIG, their predictive uncertainties (measured by average standard error) are significantly larger (28.28 and 16.89, respectively) compared to the CONFIG models (2.75 and 4.77). These large uncertainties, along with poor predictions, make such models unappealing despite their high coverage ratios. The proposed CONFIG models provide markedly better predictions with lower uncertainties while maintaining comparable coverage ratios to existing models. 

Further insight can be gleaned by comparing the performance of CONFIG with Kernel 1 vs. Kernel 2. Here, Kernel 1 provides slightly better predictions compared to Kernel 2. This is not too surprising, since this cantilever beam deflection can be viewed as a single-mechanism multi-fidelity problem and can thus be classified under Scenario 1 (see Section \ref{sec:method}), where the experiment simulates a \textit{single} mechanism with multiple fidelity parameters. Kernel 1 appears to be better suited at capturing the more complex interactions between fidelity parameters, thus leading to slightly better performance than Kernel 2.

\subsection{Quark-Gluon Plasma Evolution}
\label{sec:multistage}
We now return to our motivating problem of emulating the quark-gluon plasma, an exotic state of nuclear matter which can be created in modern particle colliders, and which pervaded the universe during its first microseconds. The study of this plasma -- in particular, properties of this unique phase of matter -- is thus an important problem in high-energy nuclear physics. Modern investigation of QGP often requires computationally intensive numerical simulations, with the plasma modeled via relativistic fluid dynamics. The use of cost-efficient emulators, when carefully constructed, can thus greatly speed up the discovery of fundamental properties on the QGP, as evidenced in recent works \citep{liyanage2022efficient,cao2021determining}.

\begin{figure}[!t]
    \centering
    \includegraphics[width=0.9\textwidth]{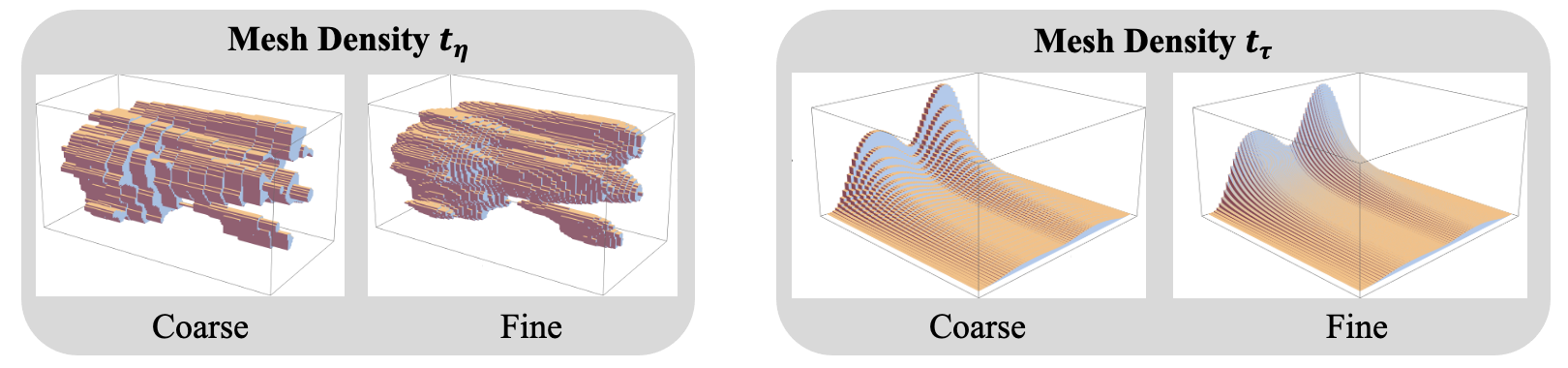}
    \caption{Visualizing the two mesh densities (fidelity parameters) in the quark-gluon plasma simulation.}
    \label{fig:QGP_mesh}
\end{figure}

For this study, we adopt a simplified version of the QGP simulation framework in \cite{everett2021multisystem}, which can be split into three distinct stages: a pre-hydrodynamic stage, a hydrodynamic stage, and a post-hydrodynamic stage. Figure \ref{fig:QGP_intro} visualizes this conglomerate (specifically, multi-stage) simulation framework. Each stage typically involves the discretization of the simulated physical system onto a spatial or space-time mesh (see Figure \ref{fig:QGP_mesh}). The sizes and dimensionalities of the meshes may vary among stages. Meshes must be large enough to contain the entire initial and final states of the systems, fine enough to capture relevant details (e.g., small-scale fluctuations in the pre-hydrodynamic initial state), and yet allow timely computation.

The considered simulator has two key fidelity parameters, $t_\eta$ and $t_\tau$, which control its precision. The first fidelity parameter arises in the pre-hydrodynamic stage. This stage models the initial distribution of energy resulting from the collision of two atomic nuclei. The energy distribution is defined on a 3D (spatial) mesh with coordinates $x$, $y$, and $\eta$.  The bounds of the mesh are fixed in all three dimensions, but the mesh density in the $\eta$ direction will be varied to adjust fidelity; it is specified by the longitudinal mesh size variable $t_\eta$, which serves as our first fidelity parameter. The simulation costs of all three stages are inversely proportional to $t_\eta$. The second parameter arises when the initial hydrodynamic state is evolved with the relativistic hydrodynamic equations until a completion criterion is reached, in effect extending the mesh into a time dimension, denoted $\tau$. The temporal mesh size variable $t_\tau$ -- our second fidelity parameter -- can thus be varied to adjust fidelity, although we note that in contrast with the $\eta$ spatial direction, the number of timesteps is not known in advance because the full evolution time cannot be fixed -- it is only determined once the completion criterion is satisfied, and so depends on the initial conditions in a complicated way. Except at very low fidelity, the simulation costs of the hydrodynamic and post-hydrodynamic stages are inversely proportional to $t_\tau$.



In this simplified QGP simulator, we consider a single response variable: the ratio of pions produced at two different points, $\eta = 0$ and $\eta = 1$. This ratio serves as a measure of how particle production is distributed along the collision axis of the atomic nuclei and is chosen because it is strongly influenced by the model parameter $\alpha$, which we use as our single input parameter in this study.
We denote $\alpha$ as $x_1$ and the ratio observable as $y_1$ below.



The experiment is carried out as follows. We compare the CONFIG models with the standard GP model and the TWY models. Since the multi-stage procedure involves multiple sequential stages, it falls under Scenario 2 (see Section \ref{sec:method}), and thus we recommend the use of Kernel 2 for CONFIG, although results for Kernel 1 are included later for completeness. We set $l = l_r = 2$ for CONFIG and $l=2$ for the two TWY models; these were optimized via cross-validation. To demonstrate the cost efficiency of multi-fidelity emulation, we again include the ``high-fidelity GP'' model, which makes use of \textit{only} high-fidelity runs to train a standard GP emulator using the squared-exponential kernel. As before, the limiting highest-fidelity setting of $\mathbf{t} = \mathbf{0}$ cannot be numerically simulated. We thus set the fidelity parameters $\mathbf{t} = (1.0 \times 10^{-4},1/64)$ as the ``high-fidelity'' setting for prediction, which appears to provide a fine enough mesh according to a mesh validation study. With this, a single high-fidelity run is very time-consuming, requiring around 1,000 CPU hours.

For comprehensive cost analysis, we fit each emulator using different design sizes, then compare the predictive performance of these models given a computational budget. The training data are generated as follows. For the high-fidelity GP model, we generate $n=2, 3, 4, \text{ or } 5$ maximin (equally-spaced) high-fidelity design points over the input interval $x_1\in[3,5]$. For the remaining models, we generate $n=15$, $20$, or $25$ design points over the joint space of input and fidelity parameters. Each design has an equal number of points on five maximin (equally-spaced) levels on $x_1$. For the two fidelity parameters $t_\tau$ and $t_\eta$, we first generate a 2D $n$-point maximin LHD \citep{morris1995exploratory} and scale this over the domain $[1.0 \times 10^{-4}, 5.0 \times 10^{-2}]\times[1/64,1/24]$. We then randomly assign to each level of $x_1$ a fidelity setting from this LHD. For validation, the test set is generated on 100 evenly-spaced points over the input space $x_1\in[3,5]$, run at the aforementioned high-fidelity setting for $\mathbf{t}$. To account for simulation variation, we repeat this procedure of training data generation and model fitting 20 times and compute the average of all metrics. The average computational cost for generating multi-fidelity training data ranges from $2.4\times 10^3$ (15 points) to $3.7\times 10^3$ (25 points) CPU hours.



\begin{figure}[!t]
    \centering
    \includegraphics[width=0.9\linewidth]{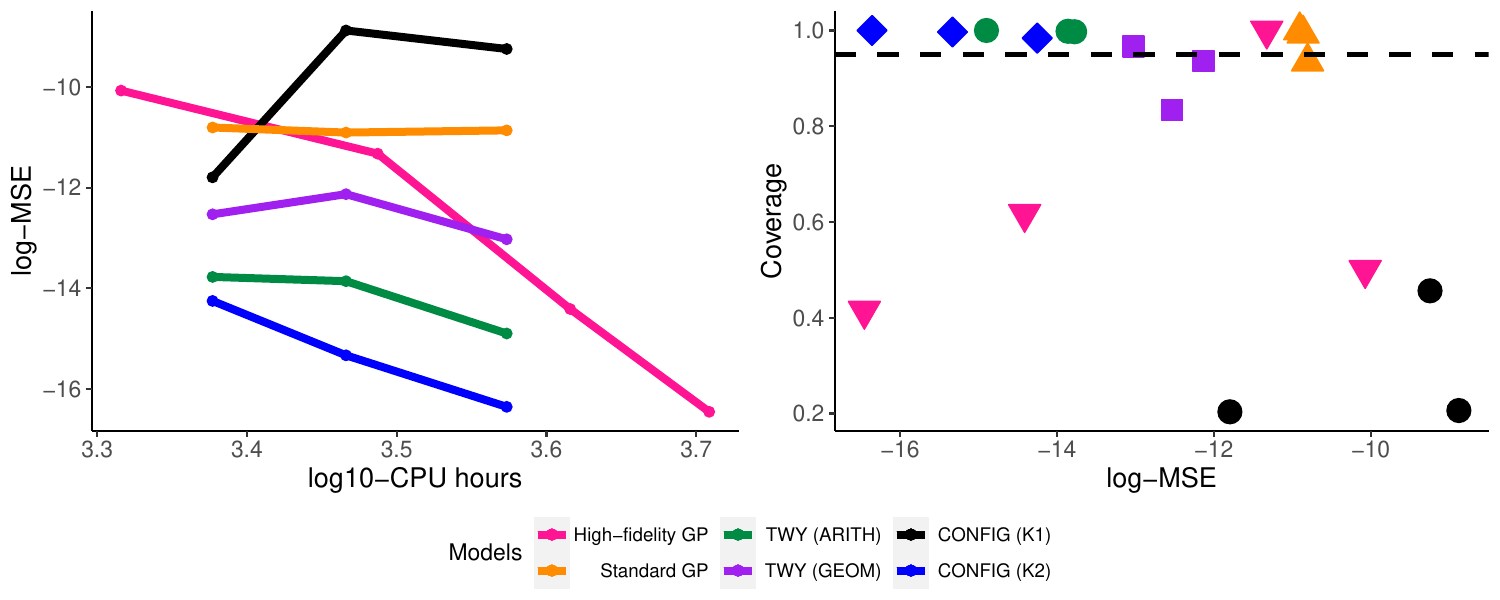}
    \caption{(Left) Plot of testing log-MSE vs. log10-CPU hours required for training data simulation for each emulator model. (Right) Scatterplot of empirical coverage ratios vs. log-MSEs for the compared models. The black dashed line denotes the nominal 95\% rate.
    }\label{fig:QGP_results}
\end{figure}



Consider first the comparison of the predictive performance of the emulators given a computational budget for training data generation. Figure \ref{fig:QGP_results} (left) plots the log-MSEs of the considered models and their corresponding costs (in log10-CPU hours) for simulating the training data. We see that, at a given computational budget, CONFIG with Kernel 2 (as recommended under Scenario 2) yields the best predictive performance out of all methods. This suggests that by integrating information on the underlying conglomerate (multi-stage) multi-fidelity simulation framework within its kernel specification, the proposed model can provide \textit{cost-efficient} and accurate emulation of expensive simulators given a tight computational budget. While such errors appear relatively small in magnitude, it is shown in \cite{TransferLearning} that small improvements in emulation accuracy may lead to large improvements (i.e., tighter constraints) for Bayesian parameter estimation of QGP properties. As such, the improved predictions from CONFIG can facilitate greater precision in scientific studies. It is also interesting to note the poor performance of CONFIG with Kernel 1 here, which we do not recommend using since this falls under Scenario 2. This is not too surprising given the fewer interactions between fidelity parameters in the current sequential multi-stage setting.

For coverage ratios, Figure \ref{fig:QGP_results} (right) shows the scatterplot of empirical coverage ratios (for 95\% predictive intervals) vs. log-MSEs for the compared methods. For the high-fidelity GP, we see that while it can achieve relatively low MSEs, it has severe undercoverage and the required cost for training data generation is high. The standard GP and the TWY models, on the other hand, provide good coverage but poor predictive performance. Of the compared models, CONFIG with the recommended Kernel 2 yields the best predictive performance with good coverage.  This again suggests that, by integrating the underlying conglomerate (multi-stage) multi-fidelity framework for non-stationary kernel specification, CONFIG can yield improved emulation performance with reliable uncertainty quantification.


\section{Conclusion}
\label{sec:conclusion}
In this paper we presented a new emulator model, called the CONFIG model, that tackles the challenge of surrogate modeling for conglomerate multi-fidelity simulators, whose precision is controlled by multiple fidelity parameters. Such simulators are often encountered in complex physical systems (including our motivating application in high-energy nuclear physics), but there has been little work in constructing cost-efficient emulators which leverage this structure for predictive modeling. CONFIG makes use of novel non-stationary covariance functions, which embed numerical convergence information on the underlying conglomerate simulator within its kernel specification. This infusion of prior information allows for effective surrogate modeling of complex simulators, even with limited training data. We demonstrate the effectiveness of the CONFIG model in a suite of simulation experiments and two applications, the first on emulating cantilever beam deflection, and the second on emulating the quark-gluon plasma in high-energy physics.


With these encouraging results, there are many avenues for future work. Given the promise of multi-fidelity modeling, one crucial direction for maximizing predictive power given a tight computational budget is experimental design. While there is a growing literature on design for multi-fidelity modeling \cite{forrester2007multi,le2015cokriging,ghoreishi2019multi,stroh2022sequential,yuchi2023design}, such methods largely do not account for multiple fidelity parameters (as is present in conglomerate simulators) or factor in varying simulation costs. For example, given a budget of $10^6$ CPU hours for a project, an experimenter would wish to know if a better predictive model can be trained with a few carefully-chosen higher-fidelity runs, or with more lower-fidelity runs. Tackling this design problem for the current conglomerate multi-fidelity framework can greatly increase the applicability of CONFIG in applications. We also aim to extend the CONFIG model for a broader range of multi-fidelity applications, where simulator fidelity is more complex and cannot be well-captured by several continuous fidelity parameters.



\if1\blind
{
\noindent \textbf{Acknowledgements}: This work was supported by DOE grants DE-FG02-05ER41367 (SAB, JFP, DS) and DE-SC0024477 (YJ, SM), and NSF grants OAC-1550225 (YJ, DS, SM), DMS-2210729 (YJ, SM) and DMS-2316012 (YJ, SM). We greatly appreciate comments and suggestions from the anonymous referees, which have improved the quality of this paper. We further thank the JETSCAPE collaboration (\url{https://jetscape.org/}) for stimulating discussions and conversations that directly motivated this work.
}
\fi
\newpage

\bibliographystyle{agsm}
\spacingset{1.05}
\bibliography{library}

\section{Appendix}
\subsection{Validation of Properties for Kernel 1}
Here, we provide a detailed explanation on how the proposed Kernel 1 satisfies the two desired properties (3.3) and (3.4). Kernel 1 takes the form:
\begin{align}
\begin{split}
    K_\mathbf{t}(\mathbf{t}_1,\mathbf{t}_2)= & \exp\left\{ -\sum_{r=1}^q \theta_r (t_{1,r}-t_{2,r})^2 \right\}
    -\text{exp}\left\{ -\sum_{r=1}^q \theta_r t_{1,r}^2 \right\}\\
    & -\text{exp}\left\{ -\sum_{r=1}^q \theta_r t_{2,r}^2 \right\}+1,
    \label{eq:kernel_1}
\end{split}
\end{align}
where $\theta_r > 0$ denotes the weight parameter for the $r$-th fidelity parameter. Let $\delta(\mathbf{t})$ denote the bias term at some fixed input $\mathbf{x}$. One way to ensure $\delta(\mathbf{t})$ satisfies the limiting condition (3.3), i.e., $\lim_{\mathbf{t} \rightarrow \mathbf{0}}\delta(\mathbf{t}) = 0$, is to represent it as a difference of two terms:
\begin{equation}
\delta(\mathbf{t}) = \kappa(\mathbf{t}) - \kappa(\mathbf{0}),
\label{eq:gul1}
\end{equation}
where $\kappa(\cdot)$ is modeled as a GP. In words, the limiting condition on $\delta(\cdot)$ is enforced by centering $\kappa$ by its response at the limiting fidelity $\mathbf{0}$. The covariance function for $\delta$ can then be written as:
\begin{align}
\begin{split}
\text{Cov}[\delta(\mathbf{t}_1), \delta(\mathbf{t}_2)] =& \,\text{Cov}[\kappa(\mathbf{t}_1)-\kappa(\mathbf{0}),\kappa(\mathbf{t}_2)-\kappa(\mathbf{0})]\\
=& \,\text{Cov}[\kappa(\mathbf{t}_1),\kappa(\mathbf{t}_2)] - \text{Cov}[\kappa(\mathbf{t}_1),\kappa(\mathbf{0})] \\
&\,- \text{Cov}[\kappa(\mathbf{t}_2),\kappa(\mathbf{0})] + \text{Cov}[\kappa(\mathbf{0}),\kappa(\mathbf{0})].
\label{eq:gul3}
\end{split}
\end{align}
Kernel 1 in \eqref{eq:kernel_1} can be recovered from \eqref{eq:gul3} with a squared-exponential correlation function on $\kappa$, and satisfies the first property (3.3) by construction as sample paths of $\kappa$ are continuous.

For the second property, note that when $t_{r'}$ approaches $0$ while $\mathbf{t}_{-r'}\neq \mathbf{0}$, we have
\begin{equation}
K_{\mathbf{t}}(\mathbf{t},\mathbf{t}) = 2 - 2\,\text{exp}\left\{ -\sum_{r=1}^q \theta_r t_r^2 \right\} = 2 - 2\,\text{exp}\left\{ -\sum_{r=1, r \neq r'}^q \theta_r t_r^2 \right\} > 0,
\end{equation}
thus satisfying the second property (3.4).

\subsection{Validation of Properties for Kernel 2}
Next, we provide a detailed explanation on how the proposed Kernel 2 satisfies the two desired properties (3.3) and (3.4). This kernel takes the form:
\begin{equation}
    K_\mathbf{t}(\mathbf{t}_1,\mathbf{t}_2) = \left[\sum_{r=1}^q \theta_r\text{min}(t_{1,r},t_{2,r})^{l_r}\right]^l,
    \label{eq:kernel_2}
\end{equation}
where $\theta_r>0$ is a weight parameter for the $r$-th fidelity parameter, and $l_r>0$ and $l>0$ are kernel hyperparameters. Note that when $\mathbf{t}=\mathbf{0}$, we have
\begin{equation}
K_{\mathbf{t}} (\mathbf{t},\mathbf{t}) = \left[ \sum_{r=1}^q \theta_r t_r^{l_r} \right]^l = \left[ \sum_{r=1}^q 0 \right]^l = 0,
\end{equation}
so $\text{Var}[\delta(\mathbf{x},\mathbf{t})]=0$. The first property (3.3) thus follows from this and the continuity of sample paths for a GP with kernel $K_\mathbf{t}$. Next, note that as $t_{r'}$ approaches $0$ when $\mathbf{t}_{-r'}\neq \mathbf{0}$, we have
\begin{equation}
K_{\mathbf{t}} (\mathbf{t},\mathbf{t}) = \left[ \sum_{r=1}^q \theta_r t_r^{l_r} \right]^l = \left[ \sum_{r=1, r \neq r'}^q \theta_r t_r^{l_r} \right]^l > 0,
\end{equation}
This suggests the variance $\text{Var}[\delta(\mathbf{x},\mathbf{t})]$ is strictly positive,  thus satisfying (3.4).

\end{document}